\documentclass[twocolumn,epjc3]{svjour3}  
\RequirePackage{graphicx}
\RequirePackage{fix-cm}
\RequirePackage{braket}
\RequirePackage{bm}
\RequirePackage{amsfonts}
\RequirePackage{amsmath,mathrsfs}
\RequirePackage{latexsym}
\RequirePackage[shortlabels]{enumitem}
\RequirePackage[numbers,sort&compress]{natbib}
\RequirePackage[colorlinks,citecolor=blue,urlcolor=blue,linkcolor=blue]{hyperref}
\RequirePackage{bbold}
\RequirePackage{dcolumn}
\RequirePackage{comment}
\allowdisplaybreaks
\newcommand{\di}{i} 
\setlist[itemize]{label=\textbullet}
\begin{document}

\title{
Multiconfigurational time-dependent density functional theory
for atomic nuclei:
Technical and numerical aspects
}

\author{Petar Marevi\'c\thanksref{e1,addr1}
        \and
        David Regnier\thanksref{addr3,addr4} 
        \and
        Denis Lacroix\thanksref{addr5}
}

\thankstext{e1}{e-mail: pmarevic@phy.hr}

\institute{Centre Borelli, ENS Paris-Saclay, 
Universit\'e Paris-Saclay, 91190 Gif-sur-Yvette, 
France \label{addr1} \and
CEA, DAM, DIF, 91297 Arpajon, France \label{addr3}
\and
Universit\'e Paris-Saclay, CEA, Laboratoire Mati\`ere en 
Conditions Extr\^emes, 91680 Bruy\`eres-le-Ch\^atel, France
\label{addr4}
\and
Universit\'e Paris-Saclay, CNRS/IN2P3, IJCLab, 91405 Orsay, France
\label{addr5}
}

\date{Received: date / Accepted: date}

\maketitle

\begin{abstract}
The nuclear time-dependent density functional theory (TDDFT)
is a tool of choice for describing various
dynamical phenomena in atomic nuclei.
In a recent study, we reported an extension of the framework 
- the multiconfigurational TDDFT (MC-TDDFT) model - that takes into
account quantum fluctuations in the collective space
by mixing several TDDFT trajectories.
In this article, we focus on technical
and numerical aspects of the model.
We outline the properties of the time-dependent
variational principle that is employed
to obtain the equation
of motion for the mixing function.
Furthermore, we discuss evaluation of various ingredients
of the equation of motion, including the
Hamiltonian kernel, norm kernel,
and kernels with explicit time derivatives.
We detail the numerical methods
for resolving the equation
of motion and outline the major
assumptions underpinning the model. A technical discussion is supplemented
with numerical examples that
consider collective quadrupole
vibrations in $^{40}$Ca, particularly focusing on the
issues of convergence, treatment of linearly
dependent bases, energy conservation,
and prescriptions for the density-dependent
part of an interaction.
\end{abstract}

\section{Introduction}
\label{intro}

The nuclear time-dependent density functional theory
(TDDFT) \cite{simenel2010,lacroix2004,simenel2012,bulgac2013,
nakatsukasa2016,stevenson2019,schunck2019}
is a tool of choice for describing the 
dynamical phenomena in atomic
nuclei such as collective 
vibrations, low-energy heavy-ion reactions, or fission. 
Similarly to TDDFT approaches
used in various branches of physics and chemistry 
\cite{casida12,marques2012}, 
it models the dynamics of a complex many-body system 
in terms of a product-type
wave function whose diabatic time
evolution is determined by a set
of Schr\"odinger-like equations
for the corresponding single-(quasi)particle states.
While such an approach includes the
one-body dissipation mechanism and is well-suited
for calculating mean values
of observables, it yields quasi-classical equations of motion in 
the collective space \cite{blaizot1986,ring2004}. Consequently,
it drastically underestimates
fluctuations of observables and
is unable to account for
quantum many-body effects such as tunneling in 
collective potential energy landscapes. 
The numerous attempts to include
quantum fluctuations beyond
the basic TDDFT framework
can be broadly classified into
two categories. 
On the one hand, the deterministic
approaches include
methods based on the
truncation of the Bogolyubov-Born-Green-Kirkwood-Yvon (BBGKY) hierarchy
\cite{bogolyubov1946,born1946,kirkwood1946}, leading to various
time-dependent $n$-particle reduced
density matrix (TD-nRDM) models 
\cite{bonitz2016,lackner2015,lackner2017},
the statistical treatment of complex
internal degrees of freedom, leading to the
Fokker-Planck framework \cite{dietrich2010},
and the Balian-V\'en\'eroni variational principle
\cite{balian1984,balian1992,simenel2011}.
On the other hand, the stochastic approaches aim to replace 
a complex initial problem with a set of simpler problems, 
as seen in the stochastic mean-field theory 
\cite{reinhard1992,reinhard1992b,lacroix2006,ayik2008,lacroix2014} 
or the exact quantum jump technique \cite{juillet2002,lacroix2005}.
However, most of these methods face challenges when applied 
in conjunction with TDDFT due to the lack of a clear 
prescription for treating effects beyond the 
independent particle approximation.
The stochastic mean-field theory, combined with TDDFT, 
has been applied with some success in describing fluctuations 
in fission \cite{tanimura2017}. Nevertheless, as demonstrated 
in Refs.~\cite{regnier2018,regnier2019}, a proper description 
of quantum effects in the collective space requires a 
genuine quantization that accounts for the interference 
between different trajectories.
Despite this, a fully quantum
multiconfigurational extension, which is nowadays routinely 
employed in static calculations
\cite{bender2003,egido2016,robledo2019,niksic2011}, 
has until very recently not been implemented in the TDDFT case.

Today, the multiconfigurational models of nuclear dynamics
are largely restricted to the adiabatic time-dependent generator coordinate
method (TDGCM) \cite{berger1984,goutte2004,younes2019,verriere2020}, 
typically supplemented with the 
Gaussian overlap approximation (GOA) \cite{brink1968,naoki1975}. 
Despite their significant success in describing numerous aspects of fission
\cite{regnier2016,schunck2016,bender2020},
the existing TDGCM implementations
consider only static states
on the adiabatic collective
potential energy landscapes 
and do not account for dissipation of the collective motion
into disordered internal single-particle motion.
Recently, the TDGCM framework was extended
with a statistical dissipative term 
in the case of fission \cite{zhao2022a,zhao2022b}, while
an earlier attempt to explicitly include two-quasiparticle 
excitations \cite{bernard2011} is yet to be implemented
in a computationally feasible framework.
However, the adiabatic approximation
still imparts significant practical and formal
difficulties to all
these models \cite{schunck2016}, including the
need to consider an extremely large number of static 
configurations, discontinuous
potential energy surfaces, and the ill-defined
scission line for fission. Consequently,
a fully quantum framework is called
for that would leverage the dissipative
and fluctuation aspects of nuclear
dynamics by removing the  adiabatic assumption and
mixing time-dependent configurations.

Theoretical foundations of such
a framework were laid out already
in the 1980s by Reinhard, Cusson,
and Goeke \cite{reinhard1983,reinhard1987}.
However, the rather limited computational capabilities of the time 
prevented any practical implementations beyond simplified models applied to
schematic problems. A step towards
more realistic applications
was recently made
in \cite{regnier2019}, where
a multiconfigurational model
was used to study the
pair transfer between two simple superfluid systems interacting 
with a pairing Hamiltonian.
Soon after, a collision between two 
$\alpha$-particles was studied within a
fully variational model based on the
Gaussian single-particle wave functions and a
schematic Hamiltonian interaction \cite{hasegawa2020}. 
While it was argued that the
model described
quantum tunneling, a discussion ensued on whether the observed phenomenon can indeed be considered tunneling
\cite{ono2022,hasegawa2022comment,ono2022reply}.
Recently, we reported the first calculations
in atomic nuclei where TDDFT configurations
were mixed based on the energy density functionals
(EDFs) framework \cite{marevic2023}.
In there, we have shown that the collective multiphonon states
emerge at high excitation energies when
quantum fluctuations in the collective space are included beyond the independent particle approximation.
A similar model, based on
relativistic EDFs, was
employed in a study of nuclear multipole vibrations
\cite{li2023} and subsequently
extended with pairing correlations to make it 
applicable to the fission phenomenon \cite{li2023b}.

The ongoing developments and the increase
of computational capabilities should, in the
near future, render these models applicable
to a wide range of nuclear phenomena.
The goal of this manuscript
is to provide more details
on technical and numerical aspects of 
the multi-configurational time-dependent
density functional theory (MC-TDDFT) framework 
reported in \cite{marevic2023}. 
In Sec.~\ref{sec:MCTDDFTState}, we outline properties
of the MC-TDDFT state and show how the time-dependent variational
principle leads
to the equation of motion
for the mixing function.
In Sec.~\ref{sec:Kernels}, we
discuss evaluation
of various ingredients
of the equation of motion,
including the Hamiltonian kernel, the norm kernel,
and kernels with explicit time derivatives. Sec.~\ref{sec:Observables}
contains details on resolving the equation
of motion and calculating various
observables, as well as a discussion
on the overall algorithmic complexity
of the framework. The technical
discussion is supplemented
with numerical examples
in Sec.~\ref{Sec:NumericalExample}. 
Finally, Sec.~\ref{sec:Conclusion} brings
summary of the present work.

\section{The MC-TDDFT state and its time evolution}
\label{sec:MCTDDFTState}
\subsection{Preliminaries}

Motivated by the well-known static generator
coordinate method (GCM) \cite{hill1953,griffin1957},
the dynamic MC-TDDFT state can be
written as
\begin{equation}
\ket{\Psi(t)} = \int\!\,d\mathbf{q}~f_{\bm{q}}(t) \ket{\Phi_{\bm{q}}(t)},
\label{eq:mctddft_state}
\end{equation}
where $\bm{q}$ 
denotes a set of continuous generating coordinates, $\ket{\Phi_{\bm{q}}(t)}$
are the time-dependent, many-body generating 
states, and $f_{\bm{q}}(t)$ is 
the mixing function which is to be
determined through a variational
principle (see Sec.~\ref{sec:eq_of_motion_f0} and \ref{sec:specificities}). 
Depending on the application,
there exists a large freedom in choosing various
ingredients of Eq.~\eqref{eq:mctddft_state}:

\begin{itemize}
\item The generating coordinates represent
collective degrees of freedom associated
to modes whose quantum fluctuations
are being considered.
In static DFT, they are often
related
to the magnitude or the phase
of a complex order parameter corresponding
to one or several broken symmetries
\cite{sheikh2019}.
Within adiabatic TDGCM studies of nuclear fission, one typically
considers multipole moments 
\cite{regnier2016,schunck2016}, 
pairing strength \cite{sadhukhan2014pairinginduced,zhao2016multidimensionallyconstrained,bernard2019role}
and occasionally
also
the nuclear temperature 
\cite{zhao2022a,zhao2022b}. 
In recent dynamical MC-TDDFT studies, 
a gauge angle was considered
as a generating coordinate
in the
case of pair transfer between
superfluid systems \cite{regnier2019},
the relative
position and momentum for collisions \cite{hasegawa2020},
and multipole boost magnitudes
or multipolarities
for vibrations \cite{marevic2023,li2023}.
\item The generating states are typically chosen as
Slater determinants \cite{hasegawa2020,marevic2023,li2023}
or the $U(1)$-symmetry-breaking
quasiparticle vacua
\cite{regnier2019}. The
corresponding single-particle wave functions
may be built upon
a simple \textit{ansatz}, such
as Gaussians
\cite{hasegawa2020}, or can
be obtained from microscopic
calculations,
based on schematic interactions
\cite{regnier2019}
or actual EDFs \cite{marevic2023,li2023,li2023b}.
In the limiting case where the generating
states are time-independent and are obtained
through energy minimization under constraints,
we recover the conventional adiabatic
TDGCM framework\footnote{In \cite{reinhard1983}, the model
based on \eqref{eq:mctddft_state} was
branded TDGCM since it represented
a time-dependent extension of the 
Hill-Wheeler-Griffin's GCM
framework \cite{hill1953,griffin1957}.
The same naming convention
was adopted in Refs. \cite{hasegawa2020}
and \cite{li2023,li2023b}.
However, over the past decade
the term TDGCM became largely associated
to adiabatic
fission models employing time-independent
generating states
\cite{younes2019,verriere2020}.
Therefore, to avoid any confusion and
underline the distinction, we 
use
MC-TDDFT to refer
to models such as the present
one that mixes states
which are not necessarily
adiabatic.}.
Irrespective of the nature of 
generating states, 
an optimal choice of the basis
set will take into account minimization
of overlaps within the set,
with the goal of reducing linear dependencies
and ensuring that each
state carries a
sufficient distinct physical information.
Any remaining linear dependencies
are later explicitly removed,
as described in Sec.~\ref{sec:InvertedNorm}
and \ref{sec:cutoff}.
\item In principle,
one could apply the variational principle with respect
to both the mixing function $f_{\bm{q}}(t)$ and the
generating states $\ket{\Phi_{\bm{q}}(t)}$. 
Such a strategy generally
yields a rather complicated set of coupled time-dependent differential equations. It is employed, for example, in quantum chemistry within
the multi-configurational time-dependent
Hartree-Fock (MC-TDHF) framework \cite{meyer2009multidimensional}.
Note, however, that this framework encompasses
only the special case 
of Hamiltonian theories with orthogonal
generating states.
Applications in
nuclear physics, where the generating
states are typically non-orthogonal,
have so far remained restricted
to the toy-model calculation
of Ref.~\cite{hasegawa2020}.
A simplification that has been adopted in recent applications \cite{regnier2019,marevic2023,li2023} is to
treat variationally
only the mixing function,
while
assuming that the generating states
follow independent trajectories.
In Ref.~\cite{reinhard1983} it was shown
that the lowest order of GOA
yields independent trajectories
even when the generating
states are treated variationally.
\end{itemize}

\subsection{Equation of motion for the mixing function}
\label{sec:eq_of_motion_f0}

The backbone idea of the MC-TDDFT framework is to look for an approximate solution of the 
time-dependent Schr\"odinger equation that takes the form of Eq.~\eqref{eq:mctddft_state}
and is parametrized
by the complex mixing function $f_{\bm{q}}(t)$.
Different variants of the time-dependent
variational principle can be used
to obtain the dynamical
equation for the mixing function \cite{yuan2019theory}.
In this work, we consider the following action:
\begin{equation}
\begin{split}
S(f,f^*,\xi_1) &= \int_{t_0}^{t_1} \,d t
\braket{\Psi(t)
| \hat{H} - \di \hbar \partial_t |
\Psi(t)} \\
&+  \int_{t_0}^{t_1} \,d t 
\, \xi_1(t) \Big(\braket{\Psi(t)|\Psi(t)} -1\Big). \\ 
\end{split}
\label{eq:augmented_action1}
\end{equation}
Here, the first term integrates the Lagrangian of the system over time and the second term imposes normalization
of the solution by
introducing a real Lagrange multiplier $\xi_1(t)$.
We look for a mixing function $f(t)$ that makes this action stationary,
\begin{equation}
\label{eq:stationary1}
    \delta S = 0,
\end{equation}
where the variation is taken with respect to any complex function $f(t)$ and any
value of the Lagrange multiplier $\xi_1(t)$,
while keeping their values fixed
at the endpoints, $\delta f(t_0)
= \delta f (t_1) = \delta \xi_1 (t_0)
= \delta \xi_1 (t_1) = 0$. 
This equation is formally equivalent to the system of equations
\begin{align}
\label{eq:stationarity_system_1}
\frac{\partial S}{\partial f} = 0, 
\quad \frac{\partial S}{\partial f^*}=0,
\quad \frac{\partial S}{\partial \xi_1}=0.
\end{align}
Writing explicitly the derivatives of the action yields a set of
integro-differential equations for the mixing function, 
\begin{subequations}
\begin{align}
\label{eq:eq_of_motion_f01}
\di \hbar \dot{f}^{\dagger} \mathcal{N}(t) 
&= f^{\dagger}(t) \Big[  -\mathcal{H}(t) +\mathcal{D}(t) - i\hbar \dot{\mathcal{N}}(t)  \nonumber \\
&  \qquad -\xi_1(t) \mathcal{N}(t)\Big], \\
\label{eq:eq_of_motion_f02}
\di \hbar \mathcal{N}(t) \dot{f}(t)
&= \Big[ \mathcal{H}(t) - \mathcal{D}(t)
+ \xi_1(t) \mathcal{N}(t)\Big]f(t),  \\
\label{eq:eq_of_motion_f03}
f^{\dagger}(t) \mathcal{N}(t) f(t) 
&= 1. 
\end{align}
\end{subequations}
We use here a compact matrix notation with respect 
to the collective coordinate $\bm{q}$.
The norm (overlap) kernel reads
\begin{equation}
\mathcal{N}_{\bm{q} \bm{q'}}(t) = 
\braket{\Phi_{\bm{q}}(t)|\Phi_{\bm{q'}}(t)}.
\label{eq:norm_kernel}
\end{equation}
Equations \eqref{eq:eq_of_motion_f01} and \eqref{eq:eq_of_motion_f02} also involve the Hamiltonian kernel
\begin{equation}
\mathcal{H}_{\bm{q} \bm{q'}}(t) = 
\braket{\Phi_{\bm{q}}(t)| \hat{H}|\Phi_{\bm{q'}}(t)},
\label{eq:h_kernel}
\end{equation}
and the time derivative kernel defined as
\begin{equation}
\mathcal{D}_{\bm{q} \bm{q'}}(t) = 
\braket{\Phi_{\bm{q}}(t)| \di \hbar \partial_t|\Phi_{\bm{q'}}(t)}.
\label{eq:d_kernel}
\end{equation}
The identity
\begin{equation}
    i\hbar \dot{\mathcal{N}}(t) = \mathcal{D}(t) - \mathcal{D}^{\dagger}(t) \label{eq:ndd}
\end{equation}
implies that Eq.~\eqref{eq:eq_of_motion_f01} is just the conjugate
transpose of Eq.~\eqref{eq:eq_of_motion_f02}. Finally, one can insert these equations into the time derivative of Eq.~\eqref{eq:eq_of_motion_f03}
to show that any value of the Lagrange parameter $\xi_1(t)$ gives a proper solution
of the system of equations as long as the initial state is normalized.
In fact, the $\xi_1(t)$ term in the equation of motion only multiplies
the function $f(t)$ by 
a time-dependent phase during the dynamics.
Setting this Lagrange parameter to zero yields the compact equation of motion
\begin{align}
\label{eq:eq_of_motion_f0}
\di \hbar \dot{f}(t)
&= \mathcal{N}^{-1}(t) \Big[ \mathcal{H}(t) - \mathcal{D}(t)
\Big]f(t).
\end{align}
This equation differs from the adiabatic TDGCM equation to the extent that
(i) all the kernels are time-dependent
and (ii) there is an additional term involving 
the time derivative kernel $\mathcal{D}(t)$.

\subsection{Dealing with time-dependent,
non-orthogonal generating states}
\label{sec:specificities}

In the present case, a particular caution is 
necessary because we employ a family of 
generating states $|\Phi_{\bm{q}}(t)\rangle$ which is
generally linearly dependent.
Consequently, the mapping between the
mixing functions $f_{\bm{q}}(t)$ and the
many-body state $\ket{\Psi(t)}$
is not bijective
\cite{ring2004}.
More specifically, at each time $t$, one may diagonalize the norm kernel 
\begin{equation}
\mathcal{N}_{\bm{q} \bm{q'}}(t) =  \sum_k \mathcal{U}_{\bm{q} k}(t) \lambda_{k}(t) \mathcal{U}_{ \bm{q'} k}^{\dagger}(t).
\label{eq:norm_diag}
\end{equation}
The columns of $\mathcal{U}(t)$ form an orthonormal
eigenbasis of the space of mixing functions $\mathcal{F}$ so that we can always
expand them as
\begin{equation}
f_{\bm{q}}(t) = \sum_k f_k(t) \mathcal{U}_{\bm{q} k}(t).
\end{equation}
Furthermore, the norm eigenvalues $\lambda_k(t)$ can be used
to split the full space $\mathcal{F}$ into the direct sum
\begin{equation}
\mathcal{F} = \mathcal{I}(t) \oplus \mathcal{K}(t).
\end{equation}
Here, $\mathcal{I}(t)$ is the \textit{image} of $\mathcal{N}(t)$, sometimes also referred to as the \textit{range} of $\mathcal{N}(t)$ \cite{saad2003iterative}. It corresponds to
the vector space of functions spanned by the 
columns of $\mathcal{U}(t)$ associated to $\lambda_{k}(t)>0$. Equivalently,
the \textit{kernel} vector space $\mathcal{K}(t)$ is spanned by the columns of $\mathcal{U}(t)$ with $\lambda_{k}(t)=0$.
In the following, we sort by convention the 
eigenvalues of $\mathcal{N}(t)$ in the ascending order so that
the first $d\!-\!r$ eigenvalues will be zero,
where $d$ and $r$ are the dimension and the
rank of $\mathcal{N}(t)$, respectively.
We can then introduce the projectors $\mathcal{P}^{\mathcal{I}}(t)$
and $\mathcal{P}^{\mathcal{K}}(t)$ on the subspaces
$\mathcal{I}(t)$ and 
$\mathcal{K}(t)$,
respectively,
with the corresponding matrices
\begin{equation}
\mathcal{P}^{\mathcal{I}}_{\bm{q} \bm{q'}}(t)
= \sum_{k > d-r} \mathcal{U}_{\bm{q} k}(t)
\mathcal{U}_{\bm{q'} k}^{\dagger}(t),
\end{equation}
\begin{equation}
\mathcal{P}^{\mathcal{K}}_{\bm{q} \bm{q'}}(t)
= \sum_{k \leq d-r} \mathcal{U}_{\bm{q} k}(t)
\mathcal{U}_{\bm{q'} k}^{\dagger}(t).
\end{equation}
The sum of the two
projectors satisfies
\begin{equation}
\mathcal{P}^{\mathcal{I}}(t) +
\mathcal{P}^{\mathcal{K}}(t) = \mathbb{1}_{\mathcal{F}}.
\label{eq:sum_of_projectors}
\end{equation}
Moreover, per definition, the norm
matrix is entirely contained in the 
image subspace and the same applies
to the kernel of any observable $\hat{O}$, 
\begin{subequations}
\begin{align}
\mathcal{P}^{\mathcal{K}}(t)\mathcal{N}(t) &= 0, \\
\mathcal{P}^{\mathcal{I}}(t)\mathcal{N}(t) &= \mathcal{N}(t), 
\label{eq:projection_N} \\
\mathcal{P}^{\mathcal{I}}(t)\mathcal{O}(t) &= \mathcal{O}(t),
\label{eq:projection_O}
\end{align}
\end{subequations}
where $\mathcal{O}_{\bm{q} \bm{q'}}(t)
= \braket{\Phi_{\bm{q}}(t)
| \hat{O} | \Phi_{\bm{q'}}(t)}$.

With these definitions at hand, we can now make 
explicit a property of the MC-TDDFT \textit{ansatz}
which is well known already from the static GCM framework~\cite{ring2004}:
at any time, the component of the mixing function $f(t)$ belonging to 
the kernel space $\mathcal{K}(t)$ does not contribute to the many-body state 
$|\psi(f(t))\rangle$. In other words,
\begin{equation}
    |\psi[f(t)]\rangle = |\psi \left[ \mathcal{P}^I(t) f(t) \right] \rangle.
\end{equation}
Going one step further, one can show that the \textit{ansatz} \eqref{eq:mctddft_state} 
provides a one-to-one mapping between the complex mixing functions
living in the space $\mathcal{I}(t)$ and the MC-TDDFT many-body states.
Although this property brings no 
formal difficulties,
it has to be taken into account
when numerically simulating the time evolution
of a system.
Indeed, the naive equation of motion \eqref{eq:eq_of_motion_f0} 
could easily lead to the accumulation
of large or even diverging components
of $f(t)$ in $\mathcal{K}(t)$, or to
fast time oscillations of $f(t)$ in this subspace.
This type of unphysical behavior may
prevent a reliable
computation of the mixing function
in practice.

To circumvent this problem, it is possible to look for 
a solution of the time-dependent variational principle that
has a vanishing component in $\mathcal{K}(t)$
for all $t$,
\begin{equation}
\label{eq:constraint_2}
 \mathcal{P}^{\mathcal{K}}(t) f(t) = 0.
\end{equation}
Such a solution is obtained by 
minimizing the augmented action 
\begin{equation}
\begin{split}
\tilde{S}(f,f^*,\xi_1,\xi_2) &= S(f,f^*,\xi_1)
\\ & + \int_{t_0}^{t_1} \,d t \,
\xi_2(t) ||\mathcal{P}^{\mathcal{K}}f||^2.
\label{eq:augmented_action}
\end{split}
\end{equation}
Compared to Eq.~\eqref{eq:augmented_action1}, this relation introduces a new term with the 
Lagrange parameter $\xi_2(t)$ that ensures the constraint \eqref{eq:constraint_2}.
The same reasoning as in Sec.~\ref{sec:eq_of_motion_f0} leads
to the modified equation of motion
\begin{equation}
\label{eq:eq_of_motion_f1}
\di \hbar \dot{f}(t) = \mathcal{N}^{-1}(t) \Big[ \mathcal{H}(t)
- \mathcal{D}(t) \Big]f(t)
+ \di \hbar \dot{\mathcal{P}}^{\mathcal{I}}(t)f(t).
\end{equation}
The last term on the right hand side 
ensures that the mixing function stays in the subspace $\mathcal{I}(t)$ at all time.
In the same way as for $\xi_1(t)$ in \eqref{eq:stationarity_system_1},
any value of $\xi_2(t)$ leads to a proper solution of the variational
principle as long as $f$ is solution of \eqref{eq:eq_of_motion_f1}.
We therefore set it to zero.
Solving Eq.~\eqref{eq:eq_of_motion_f1} 
instead of Eq.~\eqref{eq:eq_of_motion_f0} 
provides a better numerical stability at
the price of estimating
$\dot{\mathcal{P}}^{\mathcal{I}}(t)$ at each time step.

\subsection{Equation of motion for the collective wave function}
\label{sec:eq_of_motion_g}

In principle, the mixing function could be 
determined by numerically integrating
Eq. \eqref{eq:eq_of_motion_f1}. However,
like in the static
GCM,
it is useful to introduce
the collective wave function $g(t)$ as
\begin{equation}
g(t) = \mathcal{N}^{1/2}(t) f(t).
\label{eq:g_wave_function}
\end{equation}
The square root
of the norm kernel is defined by the
relation
\begin{equation}
\mathcal{N}_{\bm{q} \bm{q'}}(t) = 
\int_{\bm{q''}}
\,d\bm{q''} \mathcal{N}_{\bm{q} \bm{q''}}^{1/2}(t)
\mathcal{N}_{\bm{q''} \bm{q'}}^{1/2}(t).
\end{equation}
At any time, the collective wave function $g(t)$
belongs to the subspace $\mathcal{I}(t)$ and uniquely defines 
the MC-TDDFT state.
Following a standard procedure,
we also transform kernels
$\mathcal{O}(t)$ to their collective operators
$\mathcal{O}^c(t)$,
\begin{equation}
\mathcal{O}^c(t) = \mathcal{N}^{-1/2}(t)\mathcal{O}(t)
\mathcal{N}^{-1/2}(t).
\label{eq:collective_operators}
\end{equation}
This provides a useful mapping
\begin{equation}
\label{eq:observable}
    \langle \hat{O} \rangle(t) = g^\dagger (t) \mathcal{O}^c(t) g(t)
\end{equation}
for any observable $\hat{O}$.
Inserting the definition of the collective wave function
into Eq.~\eqref{eq:eq_of_motion_f0}
or \eqref{eq:eq_of_motion_f1} yields the equivalent equation
of motion for the collective wave function,
\begin{equation}
\di \hbar \dot{g}(t) = \Big(\mathcal{H}^c(t) - 
\mathcal{D}^{c}(t) +
\di \hbar \dot{\mathcal{N}}^{1/2}(t)\mathcal{N}^{-1/2}(t)\Big)g(t).
\label{eq:equation_g}
\end{equation}
For numerical purposes, the
total kernel on the right hand side
of Eq.~\eqref{eq:equation_g}
can be recast in an explicitly
Hermitian form,
\begin{equation}
i \hbar \dot{g}(t) = \Big[ \mathcal{H}^c(t)
+ \mathcal{T}_1^c(t) + \mathcal{T}_2^c(t) 
\Big]g(t),
\label{eq:equation_g_final}
\end{equation}
with two Hermitian kernels
\begin{subequations}
\begin{align}
\mathcal{T}_1^c(t) &= - \frac{1}{2} \big( \mathcal{D}^c(t) + \mathcal{D}^{c \dagger}(t) \big), \\ 
\mathcal{T}_2^c(t) &=  
\frac{\di \hbar}{2} \big(\dot{\mathcal{N}}^{1/2}(t)
\mathcal{N}^{-1/2}(t) -
\mathcal{N}^{-1/2}(t)\dot{\mathcal{N}}^{1/2}(t)\big).
\end{align}
\end{subequations}
The equation of motion \eqref{eq:equation_g_final}
is the one that is numerically
solved in the numerical examples
of Sec.~\ref{Sec:NumericalExample} and in our previous work \cite{marevic2023}.

\subsection{Definition of the generating states}

In this work, the generating states are built as
Slater determinants of 
independent single-particle states
\begin{equation}
\ket{\Phi_{\bm{q}}(t)} = \prod_{k=1}^{A}a_k^{\bm{q}\dagger}(t)
\ket{0},
\label{eq:slater}
\end{equation}
where $A$ is the number of particles,
$\ket{0}$ is the particle 
vacuum, and $\{a_k^{\bm{q}\dagger}(t), a_k^{\bm{q}}(t) \}$ is a
set creation and annihilation operators associated to the 
single-particle states.
The single-particle states can be
expanded in the 
spatial representation,
\begin{align}
a_k^{\bm{q}\dagger}(t) = & \sum_\sigma \int_{\bm{r}} d^3\bm{r} 
\varphi_k^{\bm{q}}(\bm{r}\sigma;t) c_{\bm{r}\sigma}^\dagger, \\
a_k^{\bm{q}}(t) = & \sum_\sigma \int_{\bm{r}} d^3\bm{r} 
\varphi_k^{\bm{q}*}(\bm{r}\sigma;t) c_{\bm{r}\sigma},
\end{align}
where $c^\dagger_{\bm{r}\sigma}$ (resp. $c_{\bm{r}\sigma}$) creates
(resp. annihilates) a nucleon of spin $\sigma$ at position $\bm{r}$.
The $k$-th single-particle wave function $\varphi_k^{\bm{q}}$ of the 
generating state labeled by $\bm{q}$ reads
\begin{equation}
\label{eq:physwf}
 \varphi_k^{\bm{q}}(\bm{r}\sigma;t) = \langle \bm{r}\sigma | a^{\bm{q}\dagger}_k (t)| 0\rangle.
\end{equation}
The single-particle wave functions \eqref{eq:physwf}
for neutrons or protons
can be decomposed as
\begin{equation}
\begin{split}
\varphi_k^{\bm{q}}(\bm{r}\sigma;t) &= \Big(\varphi_{k,0}^{\bm{q}}
(\bm{r}; t)+\di \varphi_{k,1}^{\bm{q}}(\bm{r};t)\Big)
\chi_{\uparrow}(\sigma)
\\ &+ 
\Big(\varphi_{k,2}^{\bm{q}}
(\bm{r};t)+\di \varphi_{k,3}^{\bm{q}}(\bm{r};t)\Big)
\chi_{\downarrow}(\sigma).
\label{eq:SpDecomposition}
\end{split}
\end{equation}
The four real spatial functions $\varphi_{k,\alpha}^{\bm{q}}
(\bm{r}; t)$ 
with $\alpha = 0, 3$ correspond, respectively,
to the real spin-up,
imaginary spin-up, real spin-down, and imaginary spin-down
component, and $\chi_{\uparrow/\downarrow}(\sigma)$ are the eigenstates
of the $z$ component of the spin operator.

Starting from some initial conditions, the 
generating states $\ket{\Phi_{\bm{q}}(t)}$
are then assumed to evolve independently from 
each other,
according to the nuclear
TDHF equations
\cite{simenel2010,lacroix2004,simenel2012},
\begin{equation}
\di \hbar \dot{\rho}_{\bm{q}}(t) = \Big[h[\rho_{\bm{q}}(t)],\rho_{\bm{q}}(t)\Big],
\label{eq:tdhf}
\end{equation}
where $\rho_{\bm{q}}(t)$ is the
one-body density matrix corresponding
to $\ket{\Phi_{\bm{q}}(t)}$
and $h[\rho_{\bm{q}}(t)]$ is the single-particle
Hamiltonian derived from a Skyrme EDF \cite{bender2003,schunck2019}.

\section{Calculation of kernels}
\label{sec:Kernels}

\subsection{The norm kernel}

The overlap of two
Slater determinants is given by the determinant
of the matrix containing overlaps between the corresponding
single-particle states \cite{lowdin1955,plasser2016},
\begin{equation}
\mathcal{N}_{\bm{q} \bm{q'}}(t)
= \det M_{\bm{q} \bm{q'}}(t).
\label{eq:overlap}
\end{equation}
In the absence of isospin mixing,
the total overlap corresponds to the product
of overlaps for neutrons ($\tau = n$)
and protons ($\tau = p$),
\begin{equation}
\mathcal{N}_{\bm{q} \bm{q'}}(t)
= \prod_{\tau = n, p} 
\mathcal{N}_{\bm{q} \bm{q'}}^{(\tau)}(t)= \prod_{\tau = n, p} \det M_{\bm{q} \bm{q'}}^{(\tau)}(t),
\end{equation}
where the elements of $M_{\bm{q} \bm{q'}}^{(\tau)}(t)$ read
\begin{equation}
\Big[M_{\bm{q} \bm{q'}}^{(\tau)}(t)\Big]_{kl} = 
\braket{\varphi_{k}^{\bm{q}(\tau)}(t)|\varphi_{l}^{\bm{q'}(\tau)}(t)},
\label{eq:OverlapMatrix}
\end{equation}
or explicitly
\begin{equation}
\Big[M_{\bm{q} \bm{q'}}^{(\tau)}(t)\Big]_{kl} 
= \sum_{\sigma} \int \,d^3 \bm{r}
\varphi_k^{\bm{q} (\tau) *}(\bm{r}\sigma;t) 
\varphi_l^{\bm{q'} (\tau)}(\bm{r}\sigma;t).
{\raisetag{2.0\baselineskip}}
\end{equation}

In addition to the norm kernel matrix $\mathcal{N}$(t),
Eq.~\eqref{eq:equation_g_final} requires
evaluation
of the square root of its
inverse,
$\mathcal{N}^{-1/2}(t)$.
This matrix is straightforward
to calculate when
$\mathcal{N}(t)$ is non-singular. The case of a singular
norm kernel
matrix is discussed
in Sec.~\ref{sec:InvertedNorm}.

\subsection{The Hamiltonian kernel}

\subsubsection{General expressions} 

Motivated by the generalized Wick theorem \cite{balian1969}, 
the Hamiltonian kernel\footnote{Since we are not
dealing with a genuine Hamiltonian operator
but with a density-dependent effective interaction,
the "Hamiltonian kernel" is somewhat of a misnomer.
Consequences of this distinction were thoroughly
discussed in the literature
\cite{anguiano2001,dobaczewski2007,duguet2009,robledo2010}.
The main practical
consequence for our calculations
is that it is necessary to introduce a prescription
for the density-dependent
component of an effective interaction, as explained in Sec.~\ref{sec:DensityPrescriptions}.}
can be expressed as 
\begin{equation}
\mathcal{H}_{\bm{q} \bm{q'}}(t)
= E_{\bm{q} \bm{q'}}(t) \mathcal{N}_{\bm{q} \bm{q'}}(t),
\end{equation}
where the energy kernel
$E_{\bm{q} \bm{q'}}(t)$
is obtained as a spatial
integral of the 
energy density kernel
\begin{equation}
E_{\bm{q} \bm{q'}}(t)
= \int \,d^3 \bm{r}
\mathcal{E}_{\bm{q}
\bm{q'}}(\bm{r};t).
\end{equation}
The energy density kernel
itself corresponds to the
sum of
kinetic, nuclear (Skyrme), and
Coulomb components,
\begin{equation}
\mathcal{E}_{\bm{q}
\bm{q'}}(\bm{r};t)
= \mathcal{E}_{\bm{q}
\bm{q'}}^{\rm{Kin}}
(\bm{r};t)
+ \mathcal{E}_{\bm{q}
\bm{q'}}^{\rm{Sky}}
(\bm{r};t)
+ \mathcal{E}_{\bm{q}
\bm{q'}}^{\rm{Cou}}
(\bm{r};t),
\label{eq:EnergyDensity}
\end{equation}
and it is a functional
of the one-body, non-local
transition density
\begin{equation}
\rho_{\bm{q} \bm{q'}}
(\bm{r} \sigma, \bm{r'} \sigma';t)
= \frac{\braket{\Phi_{\bm{q}}(t)|
c^\dagger_{\bm{r'}\sigma'}
c_{\bm{r}\sigma}
| \Phi_{\bm{q'}}(t)}}{\braket{\Phi_{\bm{q}}(t) |
\Phi_{\bm{q'}}(t)}}.
\label{eq:NonLocalDensity}
\end{equation}
This density
is used to derive various local
transition density
components that will appear
in \eqref{eq:EnergyDensity}.
The explicit expressions for all
the components
are given in \ref{Sec:TransitionDensities}.

\subsubsection{Energy density
components} 

To start with, the kinetic energy
density can be simply calculated
as
\begin{equation}
\mathcal{E}_{\bm{q}\bm{q'}}^{\rm{Kin}} (\bm{r};t) 
= \frac{\hbar^2}{2m} \sum_{\tau = n, p} \tau^{(\tau)}_{\bm{q}\bm{q'}}(\bm{r};t),
\end{equation}
where $m$ is the nucleon mass and 
$\tau_{\bm{q}\bm{q'}}^{(\tau)}(\bm{r};t)$ is the local
transition kinetic density [Eq.~\eqref{eq:KineticDensity}]. 

Furthermore, the 
nuclear potential component
of the energy
density is derived from the Skyrme pseudopotential
\cite{schunck2019}. The proton-neutron representation
of the energy density is equivalent to the one
used in Ref.~\cite{bonche1987},
except that the diagonal local densities
are substituted by transition local
densities defined
in \ref{Sec:TransitionDensities}.
The full expression reads
\begin{equation}
\begin{split}
\mathcal{E}_{\bm{q}\bm{q'}}^{\rm{Sky}} (\bm{r};t) &= 
B_1 \rho_{\bm{q}\bm{q'}}^2(\bm{r};t) + B_2 \sum_{\tau = n,p} \rho_{\bm{q}\bm{q'}}^{(\tau)2}(\bm{r};t) \\
& + B_3 \big(\rho_{\bm{q}\bm{q'}}(\bm{r};t) \tau_{\bm{q}\bm{q'}}(\bm{r};t) - \bm{j}_{\bm{q}\bm{q'}}^2(\bm{r};t)\big) \\ &
+ B_4\sum_{\tau = n, p}\big(\rho_{\bm{q}\bm{q'}}^{(\tau)}(\bm{r};t)\tau_{\bm{q}\bm{q'}}^{(\tau)}(\bm{r};t)
- \bm{j}_{\bm{q}\bm{q'}}^{(\tau)2}(\bm{r};t)\big) \\
& + B_5 \rho_{\bm{q}\bm{q'}}(\bm{r};t) \Delta \rho_{\bm{q}\bm{q'}}(\bm{r};t)
\\ & +  B_6 \sum_{\tau=n,p}\rho_{\bm{q}\bm{q'}}^{(\tau)}(\bm{r};t) \Delta \rho_{\bm{q}\bm{q'}}^{(\tau)}(\bm{r};t) \\
& + B_7 \rho_{D}^\alpha(\bm{r};t) \rho_{\bm{q}\bm{q'}}^2(\bm{r};t)
\\ & + B_8 \rho_D^\alpha(\bm{r};t) \sum_{\tau = n,p} \rho_{\bm{q}\bm{q'}}^{(\tau)2}(\bm{r};t) \\
& + B_9 \rho_{\bm{q}\bm{q'}}(\bm{r};t) 
\big( \nabla \cdot {\rm{\bm{J}}}_{\bm{q}\bm{q'}}(\bm{r};t) \big)
\\ & + B_9
\bm{j}_{\bm{q}\bm{q'}}(\bm{r};t) \cdot \big(\nabla \times \bm{s}_{\bm{q}\bm{q'}} (\bm{r};t)
\big)
\\
&+
B_9\sum_{\tau=n,p} 
\rho_{\bm{q}\bm{q'}}^{(\tau)}(\bm{r};t)
\big(\nabla \cdot {\rm{\bm{J}}}_{\bm{q}\bm{q'}}^{(\tau)}(\bm{r};t) \big)
\\ & + B_9
\bm{j}_{\bm{q}\bm{q'}}^{(\tau)}(\bm{r};t) \cdot 
\big(\nabla \times \bm{s}_{\bm{q}\bm{q'}}^{(\tau)}(\bm{r};t)\big) \\
& + B_{10} \bm{s}_{\bm{q}\bm{q'}}^2(\bm{r};t) + B_{11}\sum_{\tau=n,p}
\bm{s}_{\bm{q}\bm{q'}}^{(\tau)2}(\bm{r};t) \\
& + B_{12} \rho_D^\alpha(\bm{r};t) \bm{s}_{\bm{q}\bm{q'}}^2(\bm{r};t) \\ & + B_{13}\rho_D^\alpha(\bm{r};t) \sum_{\tau=n,p}
\bm{s}_{\bm{q}\bm{q'}}^{(\tau)2}(\bm{r};t). 
{\raisetag{10.0\baselineskip}}
\label{eq:skyrmekernel}
\end{split}
\end{equation}
{\vspace{10.5\baselineskip}}

\noindent The coupling constants $B_i$ and 
parameter $\alpha$ are defined in
\ref{sec:CouplingConstants}. The 
$\rho_D(\bm{r};t)$ density is defined in 
Sec.~\ref{sec:DensityPrescriptions}.

Finally, the Coulomb component is composed of the direct
and the exchange contribution,
\begin{equation}
\mathcal{E}_{\bm{q}\bm{q'}}^{\rm{Cou}} (\bm{r};t) 
= \mathcal{E}_{\bm{q}\bm{q'}}^{\rm{Cou, Dir}} (\bm{r};t) 
+ \mathcal{E}_{\bm{q}\bm{q'}}^{\rm{Cou, Exc}} (\bm{r};t).
\end{equation}
The direct contribution is calculated as
\begin{equation}
\mathcal{E}_{\bm{q}\bm{q'}}^{\rm{Cou,Dir}}(\bm{r};t)
= \frac{1}{2} ~
U_{\bm{q}\bm{q'}}^{\rm{Cou, Dir}}(\bm{r};t) \rho_{\bm{q}\bm{q'}}^{(p)}(\bm{r};t),
\end{equation}
where $\rho_{\bm{q}\bm{q'}}^{(p)}(\bm{r};t)$
is the local proton density [Eq.~\eqref{eq:ParticleDensity}]
and $U_{\bm{q}\bm{q'}}^{\rm{Cou, Dir}}(\bm{r};t)$
is the Coulomb potential
obtained as the solution
of the Poisson equation
\begin{equation}
\Delta U_{\bm{q}\bm{q'}}^{\rm{Cou, Dir}}(\bm{r};t)
= - 4 \pi \frac{e^2}{4\pi \epsilon_0} \rho_{\bm{q}\bm{q'}}^{(p)}(\bm{r};t).
\end{equation}
The real and the imaginary component
of the potential are obtained
by solving the corresponding differential
equations separately, subject to the Dirichlet
condition at the boundary
$\bm{r}_B$,
\begin{equation}
U_{\bm{q}\bm{q'}}^{\rm{Cou, Dir}}(\bm{r}_B;t) = 
\frac{e^2}{4\pi \epsilon_0}
\frac{Z_{\bm{q}\bm{q'}}(t)}{|\bm{r}_B|}.
\label{eq:cou_boundary}
\end{equation}
Here, $Z_{\bm{q}\bm{q'}}(t)$ is a complex number,
\begin{equation}
Z_{\bm{q}\bm{q'}}(t)= \int \,d^3 \bm{r} \rho_{\bm{q}\bm{q'}}^{(p)}(\bm{r};t),
\end{equation}
naturally giving a boundary condition
for both the real and the imaginary component
of the potential.
Note that the condition \eqref{eq:cou_boundary} is based 
on the multipole expansion of a generalized
charge truncated at zeroth order.
Eventually, higher orders could be included as well.
Finally, the exchage contribution is calculated
at the Slater approximation
\begin{equation}
\mathcal{E}_{\bm{q}\bm{q'}}^{\rm{Cou,Exc}} (\bm{r};t) =
- \frac{3}{4}\frac{e^2}{4 \pi \epsilon_0}
\Big(\frac{3}{\pi}\Big)^{1/3}
\Big[\rho_{D}^{(p)}(\bm{r};t)\Big]^{4/3},
\label{eq:slater_approximation}
\end{equation}
where $\rho_{D}^{(p)}(\bm{r};t)$ is the local
proton density calculated according to a prescription,
as described in Sec.~\ref{sec:DensityPrescriptions}.

\subsubsection{Density-dependent prescription}
\label{sec:DensityPrescriptions}

The local transition density is generally 
a complex
quantity and its non-integer powers are not
uniquely defined.
Consequently, a prescription is needed to evaluate
$\rho_D^{\alpha}(\bm{r})$ in \eqref{eq:skyrmekernel}
and \eqref{eq:slater_approximation}. This is a
well-known feature of multi-reference EDF models
which has been thoroughly discussed in the literature \cite{robledo2010,sheikh2019}. In the present
implementation, we opt for the average density
prescription,
\begin{equation}
\rho_D^{\alpha}(\bm{r};t) = \Big[\frac{1}{2}\Big(\rho_{\bm{q} \bm{q}}(\bm{r};t) + \rho_{\bm{q'} \bm{q'}}(\bm{r};t)\Big)\Big]^\alpha,
\label{eq:average_density1}
\end{equation}
which is always real and
reduces to the diagonal local density when
$\bm{q} = \bm{q'}$. 
An alternative form
of the average density
prescription \cite{duguet2003},
\begin{equation}
\rho_D^{\alpha}(\bm{r};t) = \frac{1}{2}\Big(\rho_{\bm{q} \bm{q}}^{\alpha}(\bm{r};t) + \rho_{\bm{q'} \bm{q'}}^\alpha(\bm{r};t)\Big),
\label{eq:average_density2}
\end{equation}
satisfies the same properties but
is obviously not equivalent to
\eqref{eq:average_density1}.
Other choices, such as the mixed density
prescription and the projected density prescription,
have also been considered in the literature,
primarily in the context of symmetry
restoration \cite{robledo2010,sheikh2019}.
Sensitivity of calculations to the
choice of prescription is discussed
in Sec.~\ref{sec:DensityPrescriptionsExample}.

\subsection{Inverse of the norm kernel}
\label{sec:InvertedNorm}

Solving Eq.~\eqref{eq:equation_g_final}
requires inverting the norm kernel matrix
$\mathcal{N}$(t). The matrix 
$\mathcal{N}^{-1/2}(t)$ is then plugged
into the last term of \eqref{eq:equation_g_final}, 
and
is also used to evaluate the collective
kernels $\mathcal{H}^c(t)$ and $\mathcal{D}^c(t)$,
according to \eqref{eq:collective_operators}.
Formally, the square root inverse $\mathcal{N}^{-1/2}(t)$
is soundly defined in the image subspace $\mathcal{I}(t)$ only.
Consequently, this linear operator always acts on functions
belonging to the image subspace,
both in the 
equation of motion [Eq.~\eqref{eq:equation_g_final}] 
and in the definition of collective kernels [Eq.~\eqref{eq:collective_operators}].
We compute its matrix elements in the $\bm{q}$ representation as
\begin{equation}
\label{eq:inverse_norm}
{\mathcal{N}}^{-1/2}_{\bm{q} \bm{q'}}(t) = 
\sum_{k > d-r} \mathcal{U}_{\bm{q} k} (t) \lambda_k^{-1/2}(t)
\mathcal{U}^{\dagger}_{\bm{q'}k}(t),
\end{equation}
where the sum runs only over strictly positive eigenvalues $\lambda_k$.

In practical applications, diagonalizing the norm kernel typically
yields several eigenvalues that are numerically close
to zero but not exactly vanishing.
It is well known from static GCM 
\cite{bender2003,robledo2019} 
and TDGCM \cite{verriere2020} that taking into account
the inverse of these eigenvalues and the associated eigenstates
in the sum \eqref{eq:inverse_norm} gives
rise to numerical
instabilities.
A standard procedure consists of introducing a cutoff parameter 
$\lambda_{\rm{cut}}$
and considering
all norm eigenvalues $\lambda_k < \lambda_{\rm{cut}}$
as numerical zeros.
In all the following applications, the square root of the inverse 
norm kernel is therefore approximated as the sum \eqref{eq:inverse_norm}
running only over the eigenvalues $\lambda_k > \lambda_{\rm{cut}}$.
The particular value
of $\lambda_{\rm{cut}}$ depends
on the application and should be 
carefully checked
on a case-by-case basis (see Sec.~\ref{sec:cutoff}).
Too large cutoff values may lead to significant errors in estimation of the inverse,
while too low values magnify numerical instabilities.
Note that the described approach is equivalent
to solving the problem in the collective 
space spanned by the so-called
natural states,
\begin{equation}
\ket{k(t)} = \sum_{\bm{q}} \frac{U_{\bm{q}k}(t)}{\sqrt{\lambda_k(t)}}\ket{\Phi_{\bm{q}}(t)},
\label{eq:CollectiveSpace}
\end{equation}
with $\dim_k \leq \dim_{\bm{q}}$, and $\dim_{\bm{q}}$
is the dimension of the $\bm{q}-$basis space.

\subsection{Kernels with explicit time derivatives}

To ensure hermiticity of the total collective kernel
on the right hand side of Eq.~\eqref{eq:equation_g_final},
it is crucial to use a consistent numerical prescription
when evaluating its various ingredients. This particularly
applies to kernels that include an explicit differentiation
with respect to time, such as the
$\mathcal{D}_{\bm{q}\bm{q'}}(t)$, 
$\dot{\mathcal{N}}_{\bm{q}\bm{q'}}(t)$, and
$\dot{\mathcal{N}}^{1/2}_{\bm{q}\bm{q'}}(t)$ kernel.

\subsubsection{The $\mathcal{D}_{\bm{q}\bm{q'}}(t)$ kernel}

We assume that
the time derivative of a generating state $\ket{\Phi_{\bm{q}}(t)}$ is
well represented by finite differences,
\begin{equation}
\partial_t \ket{\Phi_{\bm{q}}(t)} \approx
\frac{1}{\Delta t}
\big( \ket{\Phi_{\bm{q}}(t)} - \ket{\Phi_{\bm{q}}(t_-)} \big),
\label{eq:finite_differences_scheme}
\end{equation}
where $t_- = t - \Delta t$ and $\Delta t$
is the time step.
The time-derivative kernel~\eqref{eq:d_kernel}
can then be simply evaluated as
\begin{equation}
\hspace{-1mm} \mathcal{D}_{\bm{q} \bm{q'}}(t) = 
\frac{\di \hbar}{\Delta t}
\big(\braket{\Phi_{\bm{q}}(t) | 
\Phi_{\bm{q'}}(t)}
- \braket{\Phi_{\bm{q}}(t) | 
\Phi_{\bm{q'}}(t_-)} \big).
\label{eq:derivative_kernel_numerical}
\end{equation}
Calculation of the time-derivative kernel was 
therefore
reduced to evaluation of two overlap kernels
equivalent to those in
Eq.~\eqref{eq:norm_kernel}.

\subsubsection{The $\dot{\mathcal{N}}_{\bm{q}\bm{q'}}(t)$
and $\dot{\mathcal{N}}^{1/2}_{\bm{q}\bm{q'}}(t)$ kernels}

We start by evaluating the $\dot{\mathcal{N}}_{\bm{q}
\bm{q'}}(t)$ kernel, 
\begin{equation}
\dot{\mathcal{N}}_{\bm{q}
\bm{q'}}(t) =
\braket{\Phi_{\bm{q}}(t)|\dot{\Phi}_{\bm{q'}}(t)}
+ \braket{\dot{\Phi}_{\bm{q}}(t)|\Phi_{\bm{q'}}(t)}.
\end{equation}
Using the finite differences scheme
of \eqref{eq:finite_differences_scheme},
we obtain
\begin{equation}
\begin{split}
\dot{\mathcal{N}}_{\bm{q}
\bm{q'}}(t) = \frac{1}{\Delta t}
\big(
& 2 \braket{\Phi_{\bm{q}}(t)|\Phi_{\bm{q'}}(t)}
-  \braket{\Phi_{\bm{q}}(t)|\Phi_{\bm{q'}}(t_-)} 
\\ 
- & \braket{\Phi_{\bm{q}}(t_-)|\Phi_{\bm{q'}}(t)} \big).
\end{split}
\label{eq:norm_derivative_numerical}
\end{equation}
Similarly as before, we only need to evaluate three 
overlap kernels.
In the next step, we can determine the 
$\dot{\mathcal{N}}^{1/2}(t)$ kernel by
recognizing that
\begin{equation}
\dot{\mathcal{N}}(t) =
\dot{\mathcal{N}}^{1/2}(t)
\mathcal{N}^{1/2}(t)
+
\mathcal{N}^{1/2}(t)
\dot{\mathcal{N}}^{1/2}(t)
\label{eq:Ndot}
\end{equation}
represents a special
case of the Sylvester equation \cite{bhatia1997}.
If $\mathcal{N}^{1/2}(t)$ has all
positive, non-zero eigenvalues, there exists
a unique solution which can be written as
\begin{equation}
{\rm vec}\big(\dot{\mathcal{N}}^{1/2}(t)\big) = \mathcal{S}^{-1}(t) ~ 
{\rm vec}\big(\dot{\mathcal{N}}(t)\big), 
\end{equation}
where
the vectorization operator "$\rm{vec}$" corresponds
to stacking the columns of a $d \times d$
matrix into a vector of length $d^2$
and
\begin{equation}
\mathcal{S}(t) =  \mathbb{1} \otimes \mathcal{N}^{1/2}
(t)+ \big(\mathcal{N}^{1/2}(t)\big)^T \otimes \mathbb{1} 
\label{eq:sylvester_matrix_inverse}
\end{equation}
is a complex matrix belonging to $\mathbb{C}^{d^2\times d^2}$.
We recover the desired kernel in its
matrix form with the inverse of the vectorization operator.
\begin{equation}
\dot{\mathcal{N}}^{1/2}(t) =
{\rm vec}^{-1} \big[\mathcal{S}^{-1}(t) ~ {\rm vec} \big({\dot{\mathcal{N}}(t)}\big)\big].
\label{eq:n12_timederivative}
\end{equation}
Note that the described procedure requires
inverting the $\mathcal{S}(t)$ matrix, whose
dimension grows as a square of the
number of basis states. However, for the basis sizes
envisioned in applications
of MC-TDDFT (from several states to several
tens of states),
such inversions are feasible.
Should the need for even larger bases occur,
hermiticity of the 
norm matrix may be used to further
reduce the dimensionality of the problem
(for example, by using the half-vectorization instead
of the vectorization operation). 
Finally, this procedure to solve the Sylvester equation
involves inversion of the matrix $\mathcal{S}(t)$
which can be positive semi-definite, similarly to $\mathcal{N}^{-1/2}$.
Following the same procedure, 
we diagonalize the matrix
that should be inverted, keep only
the non-zero eigenvalues, and invert
the matrix in this subspace. This yields
the $\mathcal{S}(t)$
matrix which can be safely used in
Eq.~\eqref{eq:n12_timederivative}\footnote{
Several numerical tests can be performed to verify the procedure.
To start with, thus obtained $\dot{\mathcal{N}}^{1/2}(t)$
matrix should verify Eq.~\eqref{eq:Ndot}.
Moreover, it should reduce to the usual 
expression when all the
eigenvalues are non-zero.
As a third test, when plugged
into Eq.~\eqref{eq:equation_g_final},
it should lead to a unitary time
evolution.
Finally, when two identical TDDFT states
are mixed, the evolution of the MC-TDDFT
state should reduce to the evolution
of the basis state.}.

\section{Resolution of the equation
of motion and calculation of
observables}
\label{sec:Observables}

Once all the expressions
for collective
kernels
have been established
as described in 
Sec.~\ref{sec:Kernels}, 
the MC-TDDFT calculations proceed
in three major steps:
(i) choosing a set of 
initial conditions relevant for the physical case under study,
(ii) integrating in time the 
equation of motion for the basis states [Eq.~\eqref{eq:tdhf}] and the 
collective wave function [Eq.~\eqref{eq:equation_g_final}],
and (iii) computing observables of interest.

\subsection{The initial conditions}

To start with, the initial
mixing functions $f_{\bm{q}}(0)$
need to be chosen. 
This choice
is somewhat arbitrary as it
is entirely
guided by the 
physical scenario one aims
to simulate. 
For example,
in \cite{marevic2023} we 
mixed three
TDDFT states
($\bm{q} = 1, 2, 3 $) and
considered
two sets of initial conditions.
In the first case, we set
$f_1(0) = 1, f_2(0) = f_3(0) = 0$, rendering the initial
MC-TDDFT state equal to the
first TDDFT state. In the
second case, the mixing functions
were determined by diagonalizing
the initial collective Hamiltonian kernel, thus starting calculations
from the actual
multiconfigurational
ground state. Of course,
alternative choices are also
possible.

Furthermore, the initial
total collective kernel on the
right hand side of \eqref{eq:equation_g_final}
needs to be determined.
While the $\mathcal{H}^c(t)$
and $\mathcal{N}^{-1/2}(t)$
are uniquely defined at $t=0$,
this is not the case for the
$\mathcal{D}^c(t)$ and $\dot{\mathcal{N}}^{1/2}(t)$
that include explicit time
derivatives and are calculated
with the finite difference scheme.
The value of these kernels at $t=0$ is therefore estimated by
propagating the set of basis
states by $\Delta t$ and 
using the finite differences
scheme. Since for sufficiently small
time steps the collective kernels
evolve very smoothly,
the overall dynamics is not
significantly impacted by this
choice.

\subsection{Numerical
schemes for time propagation}

The nuclear TDHF equation [Eq.~\eqref{eq:tdhf}] can be efficiently
resolved by using any of the
popular numerical schemes.
In the present implementation,
we use the fourth order
Runge-Kutta method (RK4) which was in \cite{regnier2019} shown
to provide better norm conservation
properties than the Crank-Nicolson
scheme. On the other hand,
the equation
of motion for collective wave 
functions [Eq.~\eqref{eq:equation_g_final}]
is resolved by the direct method,
that is
\begin{equation}
g_{\bm{q}}(t_0 + \Delta t) = 
\exp\Big(-\frac{\di}{\hbar}\mathcal{T}(t_0) \Delta t\Big) 
g_{\bm{q}}(t_0),
\label{eq:DirectMethod}
\end{equation}
where $\mathcal{T}(t)$ is the
total collective kernel on
the right hand side of \eqref{eq:equation_g_final}.
The direct method appears
feasible for smaller
sets of basis states.
For larger sets, an alternative
method such as the RK4 may be 
better suited.

\subsection{Calculation of observables}

Following Eq.~\eqref{eq:observable}, the collective wave function can
be used to calculate
the expectation
value of any
observable in the
MC-TDDFT state at any time $t$.
Generally, the collective kernel $\mathcal{O}^c(t)$ can be calculated
from the usual kernel according to \eqref{eq:collective_operators}.
In the specific case of a one-body, spin-independent, local observable,
the generalized Wick theorem yields directly
\begin{equation}
\begin{split}
\mathcal{O}_{\bm{q} \bm{q'}}(t)  &= \mathcal{N}_{\bm{q} \bm{q'}}(t)
 \int \,d^3 \bm{r}  O(\bm{r}) 
 \rho_{\bm{q} \bm{q'}}(\bm{r};t),
\end{split}
\end{equation}
where $\rho_{\bm{q} \bm{q'}}(\bm{r};t)$ is the
transition particle density
from Eq.~\eqref{eq:ParticleDensity}
and
$O(\bm{r})$ is the
coordinate space representation
of the corresponding operator.
For example, for the multipole moment operator
$O_{lm}(\bm{r}) = r^l Y_{lm}(\theta, \phi)$, where $Y_{lm}(\theta, \phi)$
are the spherical harmonics.
Furthermore, the variance of such one-body
observable in a normalized
MC-TDDFT state can be calculated
as
\begin{equation}
\sigma^2_{\hat{O}}(t) =
\braket{\Psi(t)|\hat{O}^2|\Psi(t)}
- \braket{\Psi(t)|\hat{O}|\Psi(t)}^2.
\label{eq:variance}
\end{equation}
An explicit expression of the variance as a function of 
the one-body density is
given in \ref{sec:variance}.

\subsection{Overall algorithmic complexity}
In this section, we provide an estimate of how the 
MC-TDDFT framework scales with various parameters
of the problem.
The parameters driving the complexity of the MC-TDDFT
equation [Eq.~\eqref{eq:equation_g_final}] are the size of the spatial basis $b$
($=2 \times m_x$, where
$m_x$ is the number of mesh points
and $2$ is the spin factor), the number of particles $A$, the number of generating states $d$, and
the number of time iterations $n_t$.
For each time iteration, one needs to compute all the kernels,
propagate all the generating states,
and finally propagate the collective wave function.

Following the discussion of Sec.~\ref{sec:Kernels},
we can estimate
the computational complexity of different kernels:
\begin{itemize}
\item Norm (overlap) kernel: $O(d^2\times A^2 \times b) + O(d^2\times A^3)$.
The first term comes from the determination of all pairs of
$M_{\bm{q}\bm{q'}}$ matrices, while the second term results from the computation of their determinants. 
Since our application relies on
a spatial representation for 
which $b \gg A$, the
second term is negligible.
\item Derivative kernel:  $O(d^2\times A^2 \times b)$.
This kernel is evaluated by calculating
two overlap kernels and therefore
exhibits the same computational
complexity.
\item Hamiltonian kernel (given the overlap kernel):  $O(d^2\times A^2 \times b)$. 
Calculation of this kernel requires
inversion of all $M_{\bm{q}\bm{q'}}$ matrices at the cost of $O(d^2\times A^3)$, 
computation of all transition densities at the cost 
of $O(d^2\times A^2 \times b)$,
and finally the spatial integration of all terms in the energy density kernel at the
cost of $O(d^2 \times b)$.
Note that the zero range of the Skyrme effective interaction avoids the presence of terms proportional to $b^2$ that would significantly 
increase the computational burden.
Assuming again $b\gg A$ renders the
term associated with $d^2\times A^2 \times b$
the dominant term.
In practice, we observe that computing the Hamiltonian kernel 
is the most costly operation
(and by far more costly than the
computation of the overlap kernel,
even though they scale equally).
\item Time derivative of $\mathcal{N}^{1/2}$: $O(d^6)$. This high cost as a function of the number of 
generating states comes from the inversion of the matrix in Eq.~\eqref{eq:sylvester_matrix_inverse}.
\end{itemize}
Similar considerations yield a complexity of $O(d\times A \times b)$ for the time propagation of the generating states 
and a complexity of $O(d^2)$ for the time propagation of the collective wave function.
Keeping only the leading contributions, we conclude that the overall complexity of MC-TDDFT 
calculations using the present method scales as
$O(n_t\times d^2 \times A^2 \times b) + O(n_t\times d^6)$.
The leading contributions to this cost come from
the calculation of kernels and, more specifically, the 
calculation of transition densities (first term) 
and the inversion of the Sylvester matrix (second term).

For comparison,
a standard TDGCM model with $d'$ static generating states has a 
computational complexity of $O(d'^2\times A^2 \times b) + O( n_t\times d'^2)$. 
The first term corresponds to a single
computation of all the
time-independent kernels, while the second term corresponds to the
time propagation of the collective wave function.
Under the assumption that the two models
can describe a certain
nuclear reaction with similar
accuracy, this analysis
shows that MC-TDDFT outperforms
TDGCM 
when a small number of trajectories is sufficient.
In particular, neglecting the $O(n_t\times d^6)$ term, we estimate that
MC-TDDFT becomes computationally cheaper when
\begin{equation}
    d < \frac{d'}{\sqrt{n_t}}.
\end{equation}
In practice, the variational spaces spanned by the two
methods are different and the philosophy of 
MC-TDDFT as compared to
the standard TDGCM
is to rely on a fewer
number of generating states 
that already encompass
the time-odd physics.
Of course,
the two methods may not be equally
well-adapted to describe different
phenomena and a more quantitative comparison
dedicated to a few nuclear processes
would be welcome in the near future.

\section{Illustrative calculations}
\label{Sec:NumericalExample}

As an illustrative example,
we consider the doubly-magic
nucleus $^{40}$Ca. Like in
Ref.~\cite{marevic2023},
the calculations
are performed using a newly
developed code based on the 
finite element method
\cite{zienkiewicz2013,mfem}.
The nuclear dynamics is simulated
in a three-dimensional box
of length $L$, with a regular
mesh of $N$ cells in each
spatial direction and a finite
element basis of $n$-th order
polynomials. We employ the SLy4d
EDF \cite{kim1997}, whose
parameters were adjusted without the center of mass correction, making it particularly
well suited for dynamical studies.
Unless stated otherwise, the
average density prescription of the
form \eqref{eq:average_density1}
is used 
for the density-dependent part of 
an effective interaction.

In Sec.~\ref{Sec:NumericalExample_TDDFT},
we briefly demonstrate the convergence
of TDDFT calculations with the new code.
In Sec.~\ref{sec:CollectiveDynamics2}, we
discuss the convergence of collective dynamics
when two TDDFT trajectories are mixed.
In Sec.~\ref{sec:cutoff}, we discuss the treatment
of linear dependencies in the TDDFT basis,
using an example of mixing of three
trajectories.
Furthermore, in Sec.~\ref{Sec:EnergyConservation}
we address the issue of
energy conservation within the MC-TDDFT
framework. Finally, the influence
of the density prescription on
results is discussed in Sec.~\ref{sec:DensityPrescriptionsExample}.

\subsection{Convergence of TDDFT calculations}
\label{Sec:NumericalExample_TDDFT}

\begin{figure}
\includegraphics[width=0.49\textwidth]{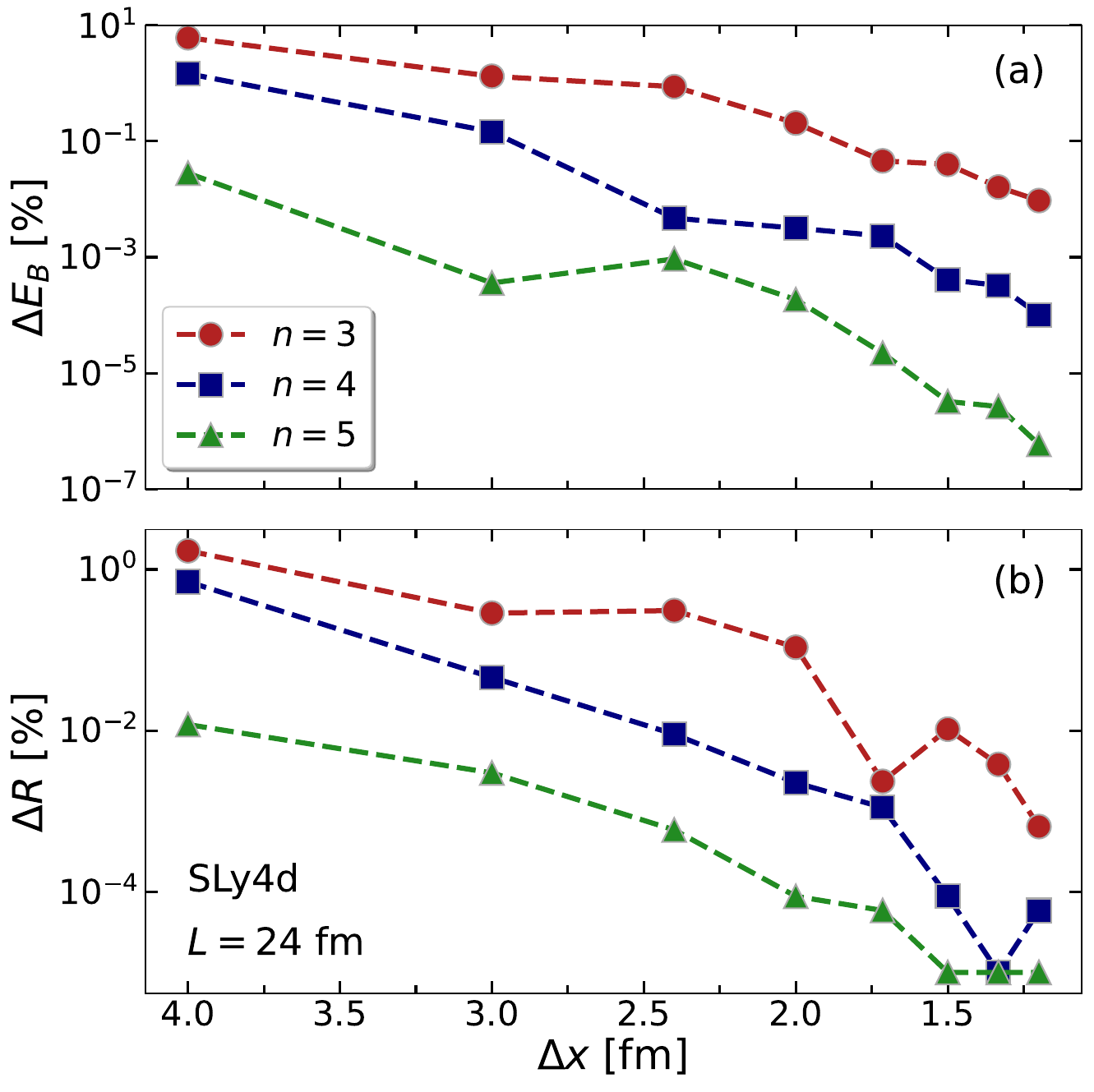}
\caption{Convergence $\Delta X = |(X - X_0)/X_0| \cdot 100$ of the ground-state binding energy ($X \equiv E_B$, panel (a)) and the ground-state
root mean square radius ($X \equiv R = \sqrt{\langle r^2 \rangle}$, panel (b)) as a function of the mesh step size
$\Delta x = L / N$, where
$L = 24$~fm is the box length
and $N$ is the number
of finite element
cells per spatial dimension.
Convergence patterns are compared
for finite element bases
of polynomials of the order $n=3,4,5$.
The fully converged values $E_0 = -339.118594$~MeV
and $R_0 = 3.413466$~fm were obtained
with $(n = 5, \Delta x = 24/26$~fm $\approx 0.92$~fm). These values agree
within at least $1$~keV and $0.003$~fm, respectively,
with those obtained using the \texttt{HFBTHO}
computational framework \cite{marevic2022}.
Note that the radius is fully converged 
for $(n = 5, \Delta x \le 1.5$~fm); the corresponding
values are plotted as $10^{-5}$.
}
\label{fig:GS_Convergence}
\end{figure}

In Fig.~\ref{fig:GS_Convergence}(a), we demonstrate the convergence of the calculated
ground-state binding energy $E_B$
by plotting $\Delta E_B = | (E_B - E_0)/E_0 | \cdot 100$
as a function of the mesh
step size $\Delta x = L/N$,
where $E_0$ is the fully converged value
(up to the sixth decimal point).
An equivalent quantity for the ground-state
root mean square radius, $\Delta R$, is shown in Fig.~\ref{fig:GS_Convergence}(b).
The box
length is fixed to $L = 24$ fm in all calculations
and polynomials of the order $n = 3, 4, 5$ are considered for the
finite element basis. 
As expected, the bases of higher
order polynomials systematically
require smaller numbers of
cells (larger $\Delta x$) to obtain a
comparable convergence. For example,
the binding energy in the ($n = 3$, $\Delta x_1 = 24/14$~fm $\approx 1.71$~fm) calculation
converges within $0.05\%$, while already
the
($n = 5, \Delta x_2 = 2.4$~fm) calculation
converges within $0.001\%$. 
The equivalent holds for radii, even 
though they converge at a somewhat
faster rate.
\begin{figure}
\includegraphics[width=0.49\textwidth]{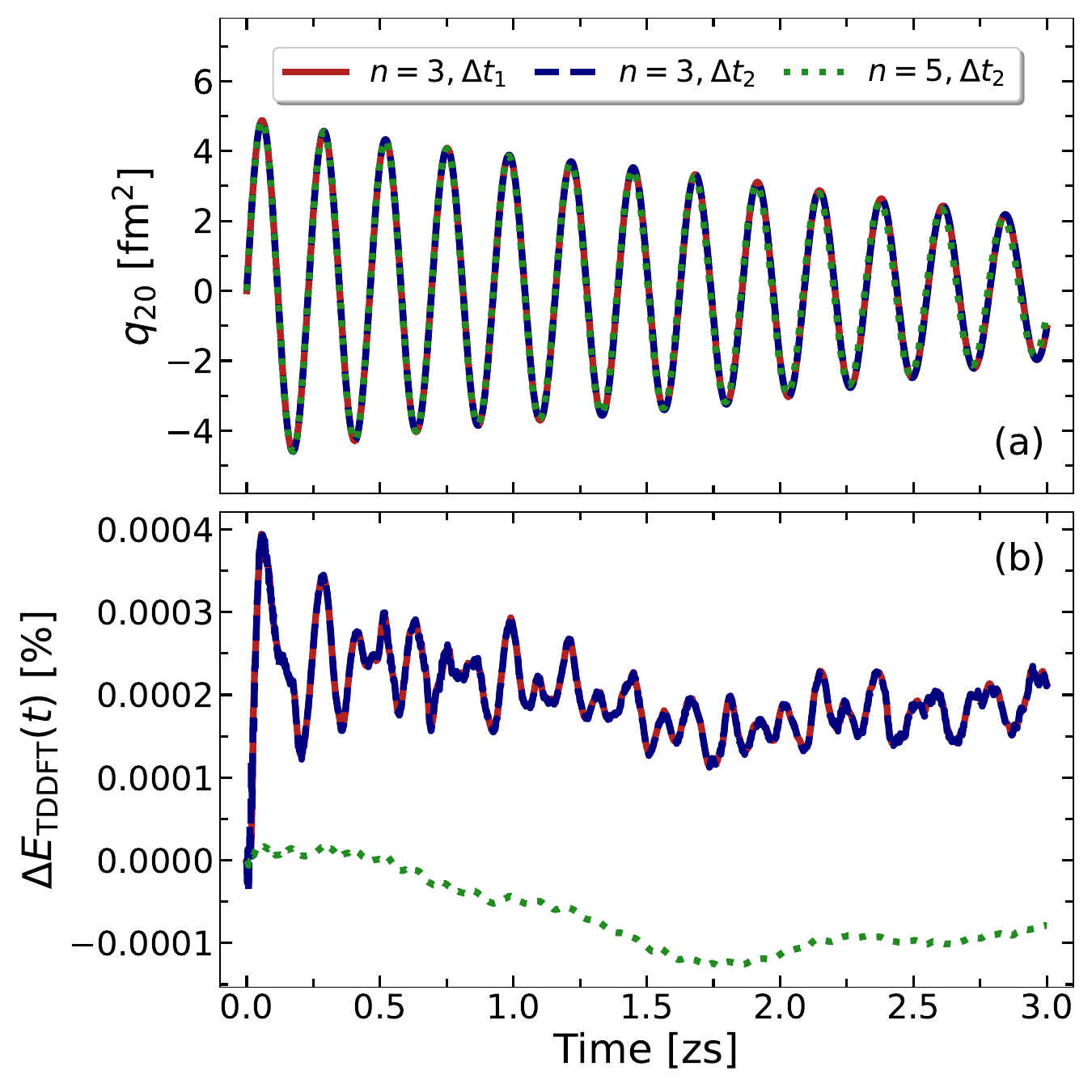}
\caption{(a): Isoscalar quadrupole moment of the
$q_{20}$-boosted TDDFT state for different
spatial and temporal discretization
schemes:
$(n=3, \Delta x_1, \Delta t_1)$
in red, $(n=3, \Delta x_1, \Delta t_2)$
in blue,
and $(n=5, \Delta x_2, \Delta t_2)$
in green,
with $\Delta x_1 \approx 1.71$~fm,
$\Delta x_2 = 2.4$~fm, $\Delta t_1 = 
5 \cdot 10^{-4}$~zs, and
$\Delta t_2 = 10^{-4}$~zs.
(b): Numerical error of the TDDFT energy,
$\Delta E_{\rm{TDDFT}}(t) = \Big(E(t) - E(0)\Big)/E(0) \cdot 100$, 
for the three cases above. 
}
\label{fig:TDDFT_Convergence}
\end{figure}

In the next step, the 
$^{40}$Ca ground state
is given an instantaneous isoscalar quadrupole boost
by applying the corresponding operator
\cite{simenel2012,scamps2013},
$\exp(\di \eta\hat{Q}_{20})$,
where $\eta = 5.7 \cdot 10^{-3}$
fm$^{-2}$ is the boost magnitude
and $\hat{Q}_{20} = r^2 Y_{20}(\theta) =
(1/4)\sqrt{5/\pi}(2z^2 - x^2 - y^2)$ is the axially-symmetric quadrupole moment operator.
To verify convergence of
the resulting TDDFT
dynamics, in Fig.~\ref{fig:TDDFT_Convergence}(a)
we compare time evolutions of the
quadrupole moment,
$q_{20}(t) = \braket{\Phi_{\bm{q}}(t)|\hat{Q}_{20}|\Phi_{\bm{q}}(t)}$,
for $(n = 3, \Delta x_1)$ and
$(n=5, \Delta x_2)$ space discretizations.
The time propagation is performed
using the fourth order Runge-Kutta 
method and is traced up to $t = 3$~zs.

The $(n = 3, \Delta x_1)$ dynamics is
well converged for a wide range
of time steps $\Delta t$; as can be 
seen in Fig.~\ref{fig:TDDFT_Convergence}(a),
the quadrupole moments $q_{20}(t)$ obtained with
$\Delta t_1 = 5 \cdot 10^{-4}$~zs
and $\Delta t_2 = 10^{-4}$~zs
are essentially indistinguishable.
A similar holds for the $(n=5, \Delta x_2)$ case,
even though achieving convergence
with higher order basis functions
and/or finer spatial meshes will
generally necessitate using
smaller time steps $\Delta t$.
For example, the $(n=5, \Delta x_2)$
calculations do not converge
for $\Delta t_1$.
However,
the $q_{20}(t)$ obtained with
$\Delta t_2$ is 
indistinguishable
from the two curves obtained with
$(n = 3, \Delta x_1)$.
This indicates that
the minor difference in ground-state
convergence of the
two sets of spatial parameters
bears no significant
consequence for the
subsequent TDDFT dynamics. 

This is further corroborated
by Fig.~\ref{fig:TDDFT_Convergence}(b),
showing the variation of the TDDFT
energy $E(t)$ as a function of time,
$\Delta E_{\rm{TDDFT}}(t) = \Big(E(t) - E(0)\Big)/E(0) \cdot 100$.
The energy should be exactly conserved within the
TDDFT framework - therefore, the small variations observed in Fig.~\ref{fig:TDDFT_Convergence}(b) stem
from numerical effects.
Standard causes for such variations include the discretization errors in the estimation of 
the spatial derivatives, leading to a non-Hermitian mean-field Hamiltonian, as well as approximations of the time propagator 
acting on the single-particle wave functions that break unitarity \cite{schuetrumpf2018tdhf,jin2021lise}.
Even though calculations with $(n=5, \Delta x_2)$ yield comparatively smaller
variations, these remain rather low
in the $(n=3, \Delta x_1)$ case; under $0.0004\%$ or less than $1$ keV. In addition,
note that variations are independent of the time step.

\subsection{Convergence of the collective dynamics}
\label{sec:CollectiveDynamics2}

\begin{figure*}[!ht]
\includegraphics[width=\textwidth]{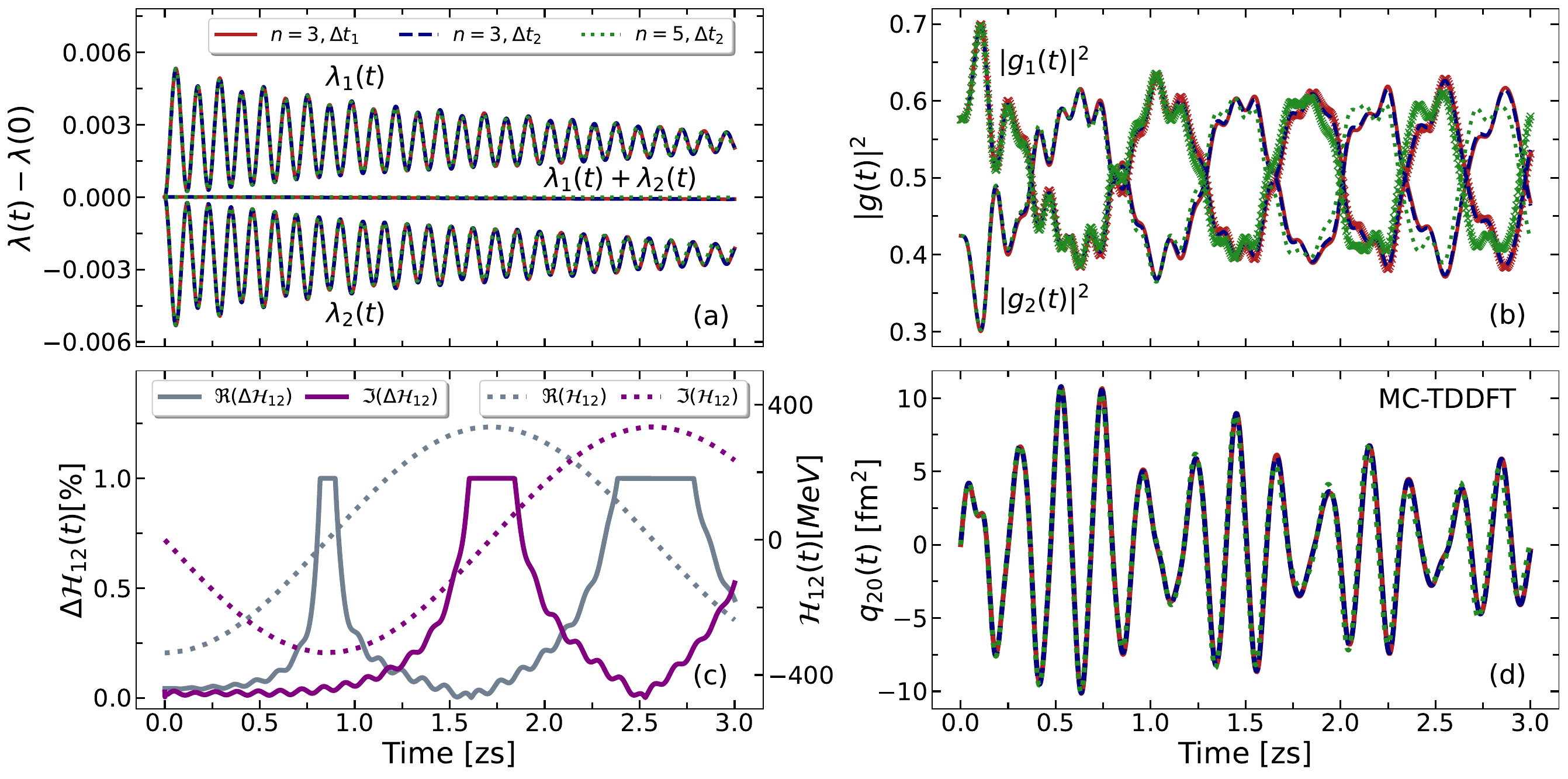}
\caption{Mixing of two TDDFT configurations, discussed in Sec.~ \ref{sec:CollectiveDynamics2}, for three
sets of spatial and temporal parameters, described
in caption to Fig.~\ref{fig:TDDFT_Convergence}.
(a): Time evolution
of the norm kernel matrix eigenvalues with respect to their initial values. 
The two
eigenvalues oscillate in counterphase,
such that their sum always remains constant. (b): Squared modulus of the collective wave function. Note that
the $|g_1(t)|^2$ component is distinguished
from the $|g_2(t)|^2$ component by the
crossed markers while keeping the same
color convention. Their sum remains equal
to one at all times, reflecting the unitarity of collective dynamics. (c): 
The difference, in percentage,
between the off-diagonal component 
of the Hamiltonian kernel
calculated with the two sets of
spatial parameters,
$\Delta \mathcal{H}_{12}(t) = \Big| \Big( \mathcal{H}_{12}^{n=3}(t) - \mathcal{H}_{12}^{n=5}(t)\Big)/\mathcal{H}_{12}^{n=5}(t) \Big| \cdot 100$,
where $\mathcal{H}_{12}^{n=3}(t)$ is calculated
with $(n=3, \Delta x_1)$ and $\mathcal{H}_{12}^{n=5}(t)$
with $(n=5, \Delta x_2)$. The right $y$-axis shows
the time evolution of $\mathcal{H}_{12}^{n=5}(t)$.
In both cases, the real and the imaginary component
are plotted separately.
The difference $\Delta \mathcal{H}_{12}(t)$ diverges
when $\mathcal{H}_{12}^{n=5}(t)$ changes sign
- consequently, all points with $\Delta \mathcal{H}_{12}(t) > 1\%$ are plotted as $1\%$.
(d): The isoscalar quadrupole moment of the 
MC-TDDFT state obtained with the three sets of parameters.
}
\label{fig:CollectiveDynamics2}
\end{figure*}

As a first example of configuration
mixing, we consider a mixed
state composed of two TDDFT configurations,
\begin{equation}
\ket{\Psi_A(t)} = f_1(t) \ket{\Phi_1(t)}
+ f_2(t) \ket{\Phi_2(t)}.
\end{equation}
Here, for $\ket{\Phi_1(t)}$ we take the TDDFT
state described in the previous section,
corresponding to the $^{40}$Ca ground
state which is given an isoscalar
quadrupole boost of magnitude $\eta_1
= 5.7 \cdot 10^{-3}$~fm$^{-2}$. Equivalently,
$\ket{\Phi_2(t)}$ corresponds to the
ground state boosted by $\eta_2 = 1.376 \cdot 10^{-2}$~fm$^{-2}$. 
This choice of quadrupole boosts yields states with excitation energies of about $E_1(t=0) = 0.25$ and $E_2(t=0) = 1.46$~MeV above the Hartree-Fock ground state.
At $t=0$,
we set $f_1(0) = 1$ and $f_2(0) = 0$
so that the initial mixed state corresponds
to the first basis state. 
The equation
of motion for the collective wave function
[Eq.~\eqref{eq:equation_g_final}]
is resolved by the direct method
[Eq.~\eqref{eq:DirectMethod}]. As before,
the box length is fixed to $L=24$~fm.
We consider the three sets of spatial
and temporal parameters described in
Sec.~\ref{Sec:NumericalExample_TDDFT}.

To start with,
the initial eigenvalues of the norm kernel matrix
read  $\lambda_1(0) = 0.011502$ and $\lambda_2(0) = 1.988498$ for the
$(n=3, \Delta x_1)$ case, 
and $\lambda_1(0) = 0.011491$ and $\lambda_2(0) = 1.988509$ and for the
$(n=5, \Delta x_2)$ case.
In Fig.~\ref{fig:CollectiveDynamics2}(a),
we show the time evolution of norm eigenvalues
with respect to their initial values,
$\lambda(t) - \lambda(0)$, for all three
sets of parameters.
The two eigenvalues oscillate in counterphase, such that their sum (the horizontal line in the middle of Fig.~\ref{fig:CollectiveDynamics2}(a)) remains
constant up to numerical accuracy.
In this case, the dimension of the collective space is the same as the
dimension of the basis space,
$\dim_k = \dim_{\bm{q}} = 2$.
However, in many practical implementations
the norm eigenstates corresponding to
very small eigenvalues will need to
be removed to ensure
a stable numerical solution.
This issue is addressed in Sec.~\ref{sec:cutoff}.

In Fig.~\ref{fig:CollectiveDynamics2}(b),
we show the squared modulus of the collective
wave function for the three
parameter sets from Sec.~\ref{Sec:NumericalExample_TDDFT}.
Once again, the $(n=3, \Delta x_1)$ calculations
are well-converged with respect to the
time step $\Delta t$. Overall,
the components of the collective wave function
exhibit an oscillatory behavior.
Their sum
remains equal
to one at all times, reflecting
the unitarity of collective dynamics.
Furthermore, the 
$(n=5, \Delta x_2)$ curves are initially
indistinguishable from the $n=3$ curves,
but start to deviate for $t > 1$~zs.

The source of these minor deviations can be traced
back to different convergence profiles of the
off-diagonal kernel elements in the
equation of motion. As demonstrated earlier in Fig.~\ref{fig:TDDFT_Convergence},
the diagonal components of the Hamiltonian
kernel - that is,
the TDDFT energies - show excellent convergence
with respect to the choice of spatial
discretization parameters.
However, the off-diagonal components involve
transition densities and are therefore
expected to exhibit weaker convergence
for the
same choice of parameters. To shed more light
on this issue,
in Fig.~\ref{fig:CollectiveDynamics2}(c) we show
the difference, in percentage,
between the off-diagonal component 
of the Hamiltonian kernel [Eq.~\eqref{eq:h_kernel}]
calculated with the two sets of
spatial parameters,
$\Delta \mathcal{H}_{12}(t) = \Big| \Big( \mathcal{H}_{12}^{n=3}(t) - \mathcal{H}_{12}^{n=5}(t)\Big)/\mathcal{H}_{12}^{n=5}(t) \Big| \cdot 100$,
where $\mathcal{H}_{12}^{n=3}(t)$ is calculated
with $(n=3, \Delta x_1)$ and $\mathcal{H}_{12}^{n=5}(t)$
with $(n=5, \Delta x_2)$.
Since $\mathcal{H}_{12}(t)$ is a complex quantity, the
corresponding real and imaginary components are 
plotted separately.
Furthermore, the right $y$-axis of 
Fig.~\ref{fig:CollectiveDynamics2}(c) shows 
the real and the imaginary component
of $\mathcal{H}^{n=5}_{12}(t)$. Please note
that this quantity is \textit{not} a constant;
the question of energy conservation is addressed 
in more detail in Sec.~\ref{Sec:EnergyConservation}.

Initially, both the real and the imaginary component of $\Delta \mathcal{H}_{12}(t)$ are
relatively small. In addition, they stay well under $1\%$
for the largest part of time evolution.
The sole exception are the regions around $t$ values
where either the real or the imaginary component
of $\mathcal{H}_{12}(t)$
changes sign (around $0.85$~zs and $2.6$~zs
for the former and $1.7$~zs for the latter).
In those cases, the denominator in $\Delta \mathcal{H}_{12}(t)$ becomes very small and
the entire quantity tends to diverge.
Consequently, for plotting purposes,
all points with $\Delta \mathcal{H}_{12}(t)
> 1\%$ are shown as $1\%$. Nevertheless, 
note that the absolute
value of deviation remains under $1$ MeV
throughout the entire time evolution.
While not drastic, such a deviation
is sufficient to cause minor
discrepancies seen in Fig.~\ref{fig:CollectiveDynamics2}(b).

To examine the impact of this effect
on an observable, in Fig.~\ref{fig:CollectiveDynamics2}(d)
we show time evolution of the isoscalar quadrupole moment 
of the MC-TDDFT state, $q_{20}(t)
= \braket{\hat{Q}_{20}}(t)$ [Eq.~\eqref{eq:observable}
with $\hat{O} = \hat{Q}_{20}$].
The three sets of parameters again
yield essentially indistinguishable results,
except
for some minor deviations
of the $(n = 5, \Delta x_2)$ curve for
larger values of $t$.

\begin{figure*}[ht]
\includegraphics[width=0.99\textwidth]{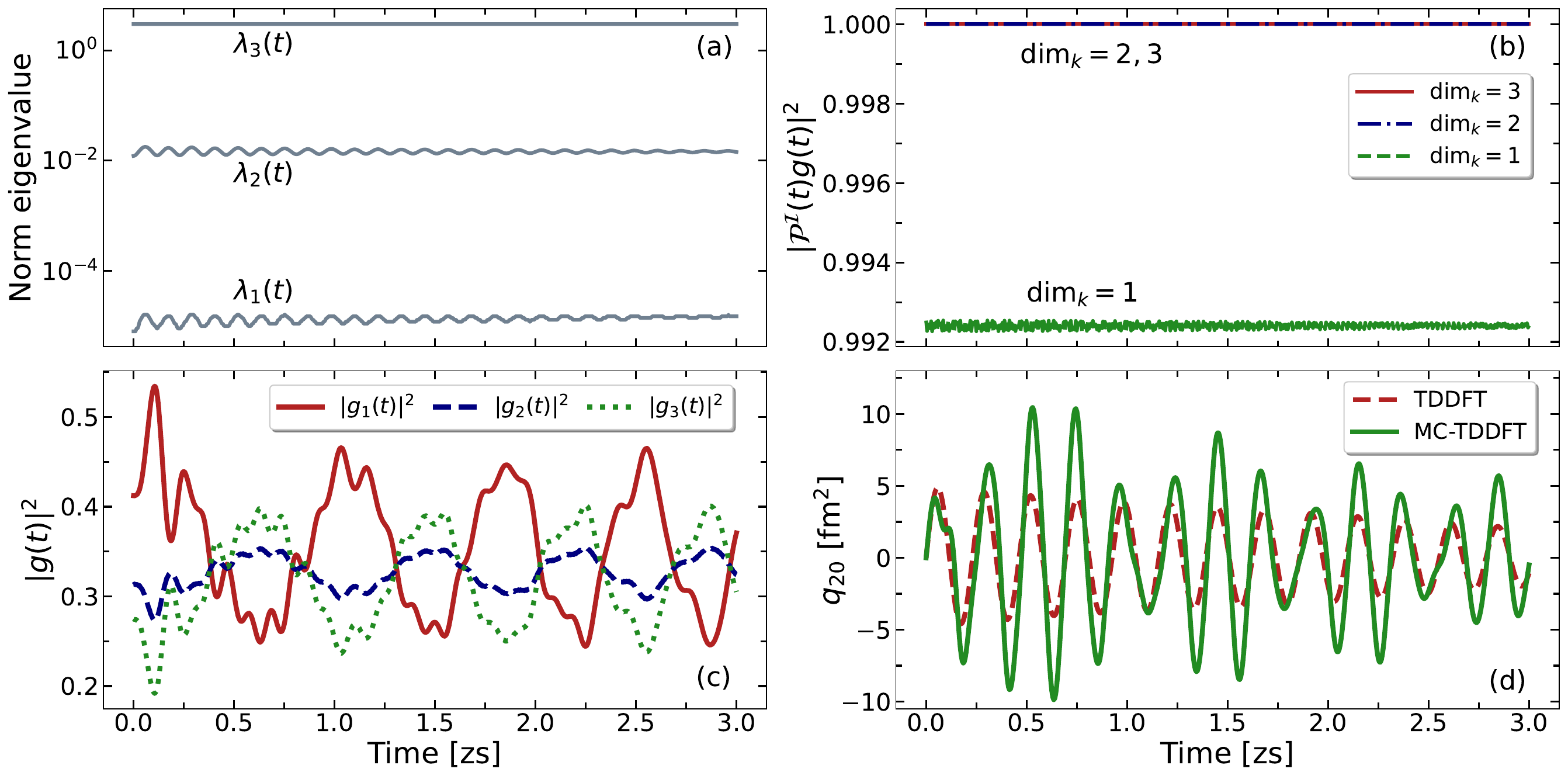}
\caption{Mixing of three TDDFT configurations, discussed in Sec.~\ref{sec:cutoff}, for 
$(n=3, \Delta x_1)$ and $\Delta t_1 = 5 \cdot 10^{-4}$~zs.
(a): Time evolution of the norm kernel
matrix eigenvalues. The three components oscillate
in time, such that their sum remains constant.
(b): Projection
of the collective wave function
onto the image subspace,
$|\mathcal{P}^{\mathcal{I}}(t) g(t)|^2$, for three different
choices of the norm eigenvalue
cutoff (or, equivalently, three
different dimensions of the collective
space $\dim_k$). The $\dim_k = 2$ provides numerical
stability while ensuring $|\mathcal{P}^{\mathcal{I}}(t) g(t)|^2 = 1$ up to $\approx 10^{-7}$.
(c): Squared modulus of the collective wave function
for $\dim_k = 2$.
(d): The isoscalar quadrupole moment
of a single TDDFT trajectory (red) and the MC-TDDFT
state (green) for $\dim_k = 2$.
}
\label{fig:CollectiveDynamics3}
\end{figure*}

Overall, a particular choice of spatial
parameters will reflect a compromise
between the feasibility of computational cost and
the required accuracy. 
In the current case, the
$(n=5, \Delta x_2)$ parameters appear to yield
somewhat more accurate results,
but at the price of about three times
longer computational time per
iteration. Of course, this difference
will become even larger as the
basis size increases.
Therefore, in the following, we will be using
the $(n=3, \Delta x_1)$ spatial
parametrization - a choice which
was also made in Ref.~\cite{marevic2023}.

\subsection{Treatment of linear dependencies
in the basis}
\label{sec:cutoff}

In the next example, we consider
mixing of three TDDFT configurations,
\begin{equation}
\ket{\Psi_B(t)} = \sum_{i=1}^3 f_i(t) \ket{\Phi_i(t)}.
\end{equation}
Here, $\ket{\Phi_1(t)}$
and $\ket{\Phi_3(t)}$
correspond to the
two configurations from the previous section.
Furthermore, $\ket{\Phi_2(t)}$ is generated
in the same manner, by 
applying the $q_{20}$-boost
of the magnitude 
$\eta_2 = 1.14 \cdot 10^{-2}$~
fm$^{-2}$ to the ground
state. This yields
an excited state with  $E_2(t=0)=1.00$~MeV. 
This choice of 
basis states is equivalent to the
one made in Ref.~\cite{marevic2023}.
We also set $f_1(0) = 1$ and $f_2(0) = f_3(0) = 0$ so that
the initial mixed state corresponds
to the first basis state and coupling
between the trajectories kicks in
with time. We use $(n=3, \Delta x_1)$,
$\Delta t_1 = 5
\cdot 10^{-4}$~zs,
and set the box size 
to $L=24$~fm.

In Fig.~\ref{fig:CollectiveDynamics3}(a), the time evolution
of the eigenvalues of the 
corresponding norm matrix are shown.
Starting from $\lambda_1(0) = 8 \cdot 10^{-6}$, $\lambda_2(0) = 0.012162$, and
$\lambda_3(0) = 2.987830$,
the three components oscillate
in time, such that their sum remains 
constant (up to numerical accuracy).
The amplitude of these oscillations
is rather small, in agreement
with the fact that each basis
state itself describes a small-amplitude
nuclear oscillation.

The collective wave function should, at all times,
be contained in the image of $\mathcal{N}(t)$.
In Fig.~\ref{fig:CollectiveDynamics3}(b),
we show the projection of the collective wave function
onto the image subspace, $|\mathcal{P}^{\mathcal{I}}(t)g(t)|^2$, for three different choices of the
collective space [Eq.~\eqref{eq:CollectiveSpace}]. 
For $\dim_k = \dim_{\bm{q}} = 3$, per definition,
we have $|\mathcal{P}^{\mathcal{I}}(t)g(t)|^2 = 1$ for all $t$. However, the
inclusion of a very small eigenvalue $\lambda_1(t)$
causes numerical instabilities. This leads to,
for example, spurious small oscillations
of the center of mass or to the
total collective kernel on the right hand side
of Eq.~\eqref{eq:equation_g} not being exactly Hermitian.
On the other hand, the $\dim_k = 1$ choice 
yields $|\mathcal{P}^{\mathcal{I}}(t)g(t)|^2 ~\approx 0.992$, reflecting the fact that removing the 
relatively large 
norm eigenstate with $\lambda_2(t) \approx 0.01$
removes a portion of physical information as well. 
The collective space should correspond to the smallest
subspace of the full Hilbert space
containing all the basis states;
for too large cutoffs, however, the collective space
does not anymore contain all the basis states.
Consequently, the $\dim_k = 2$ choice is optimal in this case -
while being numerically stable, it also ensures
$|\mathcal{P}^{\mathcal{I}}(t)g(t)|^2 = 1$
up to $\approx 10^{-7}$.
An equivalent analysis could be carried out
looking at the Frobenius norm of 
$|\mathcal{P}^{I}(t)\mathcal{N}(t) - \mathcal{N}(t)|$
[Eq.~\eqref{eq:projection_N}]
or to the partial sum
of eigenvalues $\sum_{k \in \rm{Im}(\mathcal{N})} \lambda_k$.

In Fig.~\ref{fig:CollectiveDynamics3}(c), we 
show the squared modulus of the resulting collective
wave function, which again
exhibits an oscillatory behavior. 
In particular, one can notice a close
resemblance of $|g_1(t)|^2$ and $|g_3(t)|^2$ to the collective
wave function from Fig.~\ref{fig:CollectiveDynamics2}(b).
This is entirely expected, since the corresponding
collective spaces are both of dimension $2$ and
spanned by very similar states.

Finally, Fig.~\ref{fig:CollectiveDynamics3}(d) shows
the isoscalar quadrupole moment of
the $\ket{\Phi_1(t)}$ TDDFT trajectory and 
of the MC-TDDFT state. As already noted in
Ref.~\cite{marevic2023}, the TDDFT curve
exhibits nearly harmonic oscillations of a single frequency, consistently with what is usually observed within TDDFT when the Landau damping effect is absent.
The other two TDDFT trajectories (not shown) oscillate
at the same frequency, but with slightly
larger amplitudes.
On the other hand, the MC-TDDFT curve is markedly more complex, exhibiting multiple frequencies, which
can be related to the emergence of collective 
multiphonon excitations
\cite{chomaz1995,aumann1998} in a requantized
collective model. 
Consistently with remarks above,
the MC-TDDFT curve closely resembles the corresponding
curves from Fig.~\ref{fig:CollectiveDynamics2}(d).
Calculating the Fourier transform of the quadrupole
response yields the multiphonon spectrum,
as discussed in Ref.~\cite{marevic2023}
and Sec.~\ref{sec:DensityPrescriptionsExample}.
The MC-TDDFT model is marked by a significantly larger fluctuation in quadrupole
moment - see Ref.~\cite{marevic2023} for physical
discussion and \ref{sec:variance} for technical
details of calculating it.

\subsection{Conservation of energy}
\label{Sec:EnergyConservation}

\begin{figure}
\includegraphics[width=0.49\textwidth]{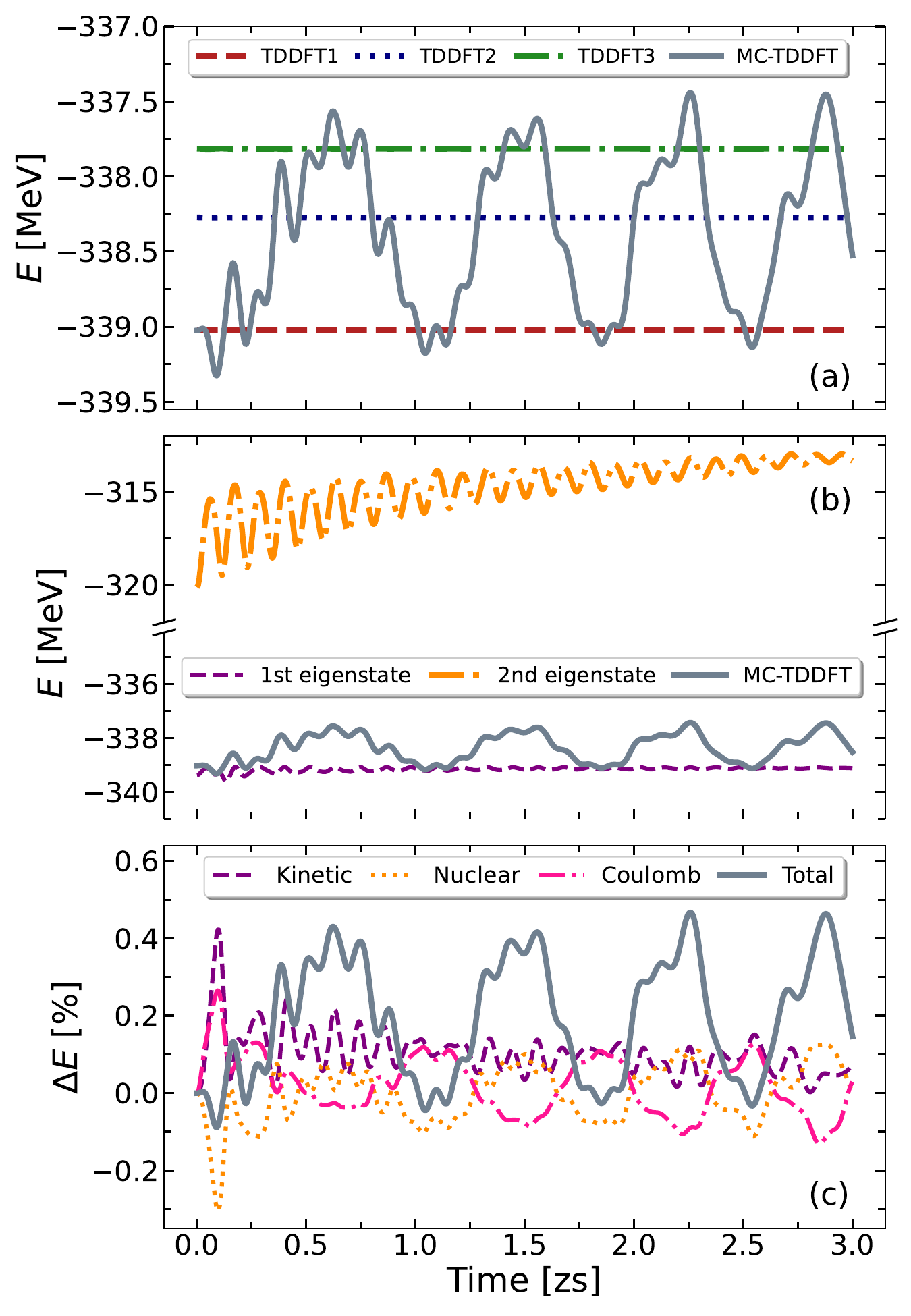}
\caption{Energy conservation in the
case of mixing of three TDDFT configurations, discussed
in Sec.~\ref{sec:cutoff},
for
$(n=3, \Delta x_1)$ and $\Delta t_1$.
(a): Time evolution of energies of the TDDFT configurations
and of the MC-TDDFT state. While
TDDFT energies are constants
of motion, the MC-TDDFT
energy is not conserved. (b) Time evolution of the 
eigenvalues of the collective Hamiltonian and of the
MC-TDDFT state. Note the break along the $y$-axis. The MC-TDDFT energy is bounded
by the two eigenvalues.
(c): Deviation, in percentage, of different components
of the MC-TDDFT energy, $\Delta E_i = \Big( E_i(t) - E_i(0) \Big) / |E_i(0)| \cdot 100$. The kinetic,
nuclear (Skyrme), and Coulomb component
all contribute to the variation of the 
total energy.
}
\label{fig:EnergyConservation}
\end{figure}

\begin{figure*}
\includegraphics[width=0.99\textwidth]{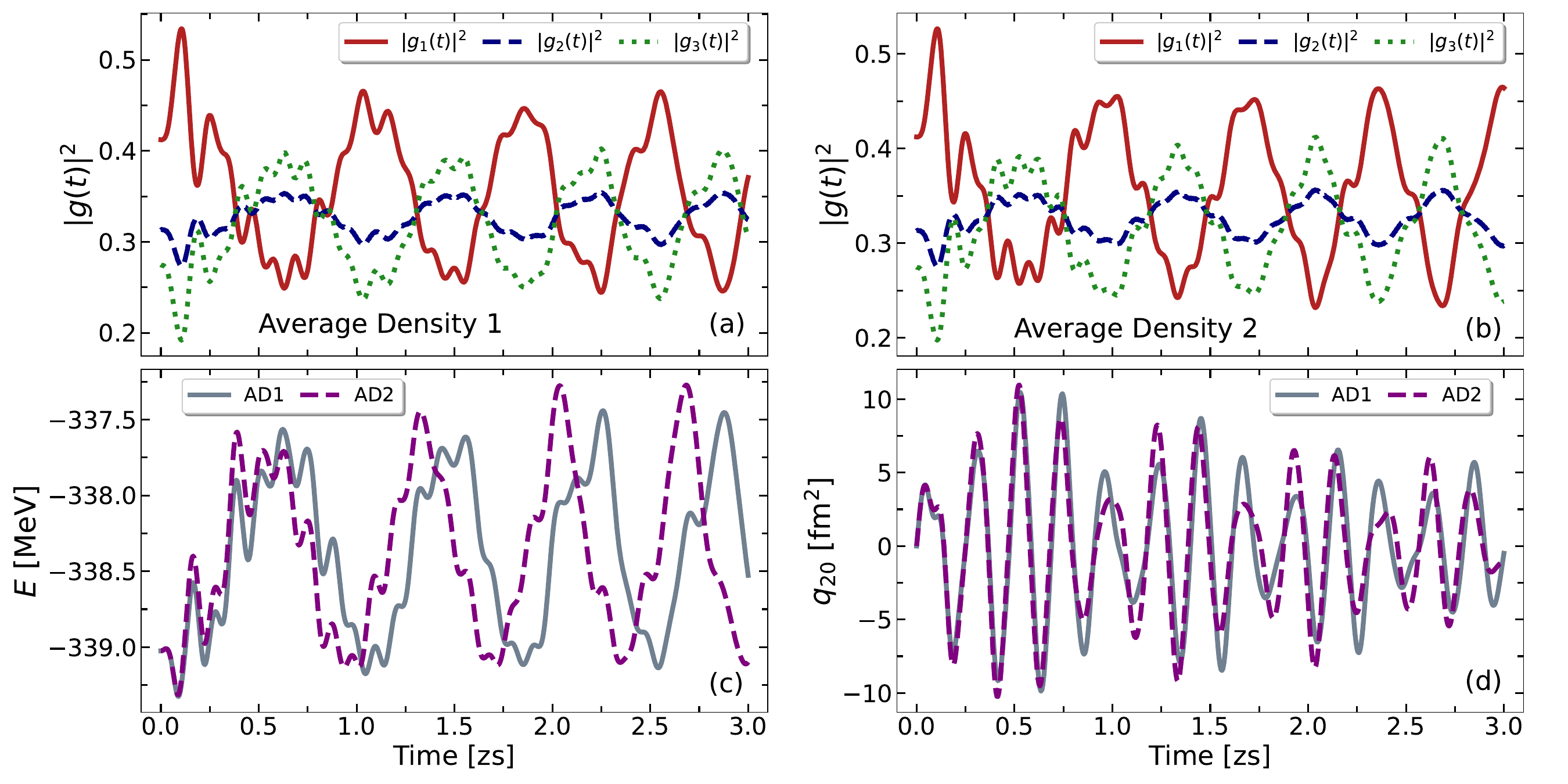}
\caption{Influence of the choice of the density-dependent
prescription in the case of mixing of three TDDFT configurations,
discussed in Sec.~\ref{sec:cutoff},
for $(n=3, \Delta x_1)$ and $\Delta t_1$.
(a): Squared modulus of the collective wave function
with the AD1 prescription
[Eq.~\eqref{eq:average_density1}].
(b): Squared modulus of the collective wave function
with the AD2 prescription
[Eq.~\eqref{eq:average_density2}].
(c): Comparison of the MC-TDDFT energy obtained with 
the two prescriptions.
(d): Comparison of the isoscalar quadrupole moment
of the MC-TDDFT state obtained with the two prescriptions.
}
\label{fig:DensityPrescription}
\end{figure*}

To address the question of energy conservation within the MC-TDDFT framework, we
consider the same example of
mixing of three TDDFT configurations
from the previous section.
In Fig.~\ref{fig:EnergyConservation}(a),
we show the time evolution
of energies of the three TDDFT trajectories,
as well as the energy of the MC-TDDFT state.
As expected, the TDDFT energies are constant
(up to numerical accuracy, see discussion in  Sec.~\ref{Sec:NumericalExample_TDDFT}). On the other hand,
the MC-TDDFT energy is markedly \textit{not} a constant.

Due to the choice of initial conditions, at $t=0$ this energy corresponds
to the energy of the first basis state. However, for $t>0$,
the MC-TDDFT energy oscillates with the amplitude
of about $1.5$ MeV. What may seem surprising at first glance
is that these oscillations are not bounded by the
TDDFT energies. In fact, it is straightforward to show that
the amplitude of oscillations
is limited by the
eigenvalues of the collective Hamiltonian and not by the
energies of the basis states.

Following Eq.~\eqref{eq:observable},
the MC-TDDFT energy can be calculated
as
\begin{equation}
E_{\rm{MC-TDDFT}}(t) = g^\dagger (t)
\mathcal{H}^c(t) g(t).
\end{equation}
The collective Hamiltonian $\mathcal{H}^c(t)$
can be recast into the diagonal form,
\begin{equation}
\mathcal{H}^c(t) = U_H(t) \Lambda_H(t) U_H^\dagger(t).
\end{equation}
Since the collective Hamiltonian matrix is Hermitian,
$\Lambda_H(t)$
is a real and diagonal matrix of eigenvalues
and $U_H(t)$ is the unitary
matrix of eigenstates.
Consequently,
we have
\begin{equation}
E_{\rm{MC-TDDFT}}(t) = \tilde{g}^\dagger(t)
\Lambda_H(t) \tilde{g}(t),
\end{equation}
where $\tilde{g}(t) = U_H^\dagger(t) g(t) U_H(t)$. 
It then follows that the MC-TDDFT
energy is bounded by the lowest and the highest eigenvalue
of the collective Hamiltonian.

In Fig.~\ref{fig:EnergyConservation}(b),
we show the time evolution of the eigenvalues
of the collective Hamiltonian
($\dim_k = 2$), alongside
the energy of the mixed state.
We verify that the MC-TDDFT energy is indeed
bounded from below by the lowest
eigenvalue of the collective Hamiltonian.
Moreover, it never even
approaches the upper limit, which
is given
by the second eigenvalue. This is consistent
with the fact that the dynamics of the
mixed state is largely
driven by the dominant eigenstate of the
collective Hamiltonian, with only minor
admixtures of the other eigenstate.

To gain more insight into the
energy of the mixed state,
in Fig.~\ref{fig:EnergyConservation}(c)
we show variations, in percentage, of different
components of the MC-TDDFT energy. More precisely,
we calculate $\Delta E_i = \Big( E_i(t) - E_i(0) \Big) / | E_i(0)| \cdot 100$, for $i = $ kinetic, nuclear (Skyrme), Coulomb.
It is apparent that neither component alone is responsible
for the variation of the total energy. Rather than that,
variations in all components fluctuate with the
magnitude of $<0.5\%$. It is worth
noting that this implies that the variations 
are unlikely to stem
from any issue related to the density-dependent
prescription. Indeed, the kinetic energy, 
which is free from
any such spuriosities, nevertheless
exhibits comparable variations.

Actually, the non-conservation of
energy appears to be
a feature of multiconfigurational models
that are not fully variational
\cite{regnier2019,marevic2023,li2023,li2023b}. 
These models are formulated under
a simplifying assumption that
the time evolution of each TDDFT
trajectory can be performed
independently, using the existing
TDDFT solvers. While
significantly alleviating the computational
burden, such 
an approximation disregards the feedback between the evolution
of the mixing function and the 
basis states,
thus removing the mechanism
that would enforce a strict
energy conservation on the
MC-TDDFT level. At best, the
approximate energy conservation
can be imposed by controlling
the lower and the upper limit of the
collective Hamiltonian eigenvalues.
This is a clear
shortcoming of such models,
which can be understood as an intermediate
step that renders the computational
implementation feasible and enables
pioneering exploration
of the MC-TDDFT capabilities in atomic nuclei.
Extending the model with a variational principle
that treats both the mixing function
and basis states as variational
parameters, like in the toy model
study of 
Ref.~\cite{hasegawa2020},
is a natural method of ensuring
the full energy conservation, at
the price of explicitly including
the effect of configuration mixing
on individual TDDFT trajectories.

\subsection{The density-dependent
prescription}
\label{sec:DensityPrescriptionsExample}

As mentioned in Sec.~\ref{sec:DensityPrescriptions}, a prescription is needed to evaluate
the $\rho_D^\alpha(\bm{r})$ density
in \eqref{eq:skyrmekernel}
and \eqref{eq:slater_approximation}.
To quantify the impact of this choice
on the nuclear dynamics, we adopt the
same example as in the previous
two subsections. This time, however,
we consider two different prescriptions
for the density-dependent part of an
effective interaction. In addition to the average density prescription of the form \eqref{eq:average_density1}, which
was used in all the calculations
up to now, we also consider the prescription
of the form \eqref{eq:average_density2}.
Both prescriptions yield real densities
and reduce to the diagonal
local density in the TDDFT limit.
However, they are evidently not equivalent
and some dependence of nuclear dynamics
on this choice is, therefore, expected.

In Figs.~\ref{fig:DensityPrescription}(a)
and \ref{fig:DensityPrescription}(b),
we compare the collective wave function
obtained with the prescription 
\eqref{eq:average_density1} ("Average Density 1" - AD1) and the prescription
\eqref{eq:average_density2} ("Average Density 2" - AD2). The main difference between
the two panels appears to be a moderate
shift in phase for larger
values of $t$. Other than that,
the overall dynamics seems relatively
unaffected. 

The impact on observables is examined
in Figs.~\ref{fig:DensityPrescription}(c)
and \ref{fig:DensityPrescription}(d),
where we compare the energy and
the isoscalar
quadrupole moment of the 
MC-TDDFT state, respectively,
obtained with the two prescriptions.
The amplitude of variations in energy 
remains very similar
for the two prescriptions.
On the other hand, there appears
a moderate shift in phase,
similar to the one seen in the collective
wave function.
Furthermore, the quadrupole moment is essentially
unaffected up to $t \approx 0.75$~zs.
Beyond this point, the difference
in prescriptions starts to play a role,
causing a moderate difference between
the two curves. 

One way to additionally quantify this difference
is to calculate the excitation spectrum
by performing a Fourier transform
of the quadrupole response.
In \cite{marevic2023}, we calculated
this spectrum using the 
AD1 prescription
and obtained the main giant resonance
peak at about
$18$ MeV, in agreement with
experiments. Moreover, two additional
peaks were observed at
approximately twice and thrice the energy
of the main peak, which were interpreted
as multiphonon excitations of the
main giant resonance. By repeating the
same procedure for the response
obtained with the AD2 prescription,
we obtain essentially the same spectrum,
with all the peaks shifted by less
than $0.1$~MeV. This is an encouraging 
result, indicating that the main conclusions
of \cite{marevic2023} may not be very sensible
to the choice of density prescription.

\section{Summary}
\label{sec:Conclusion}

The nuclear TDDFT framework is a tool of choice
for describing various dynamical phenomena in atomic
nuclei. However, it yields quasi-classical equations
of motions in the collective space and, consequently,
drastically understimates fluctuations of observables.
On the other hand, the MC-TDDFT model
encompasses both
the dissipation and 
the quantum
fluctuation
aspects of nuclear dynamics
within a fully quantum
framework. Starting from
a general mixing of diabatic many-body configurations,
the time-dependent variational principle
yields the equation of motion for the mixing
function whose resolution provides
an access to various observables of interest.

In Ref.~\cite{marevic2023}, we reported a study of
quadrupole oscillations in $^{40}$Ca where several 
TDDFT configurations were mixed based on the
Skyrme EDF framework. We demonstrated that the collective
multiphonon states emerge at high excitation energies
when quantum fluctuations in the collective space
are included beyond the independent particle
approximation. In this work, we provided more
technical and numerical details of the underlying 
MC-TDDFT model.

The central equation of motion [Eq.~\eqref{eq:equation_g_final}], obtained through
a time-dependent variational principle with the mixing
function as a variational parameter, describes
a unitary time evolution of the collective wave function.
We discussed methods for consistent computation of different
ingredients of the equation, including the Hamiltonian kernel,
norm kernel, and kernels with explicit time derivatives,
as well as the choice of initial conditions and the
direct resolution method. A special attention
needs to be given when inverting the norm kernel matrix,
since linear dependencies in the TDDFT basis can lead
to numerical instabilities. 
Within the current implementation
of the model, 
the TDDFT
configurations are assumed to evolve independently. This 
approximation simplifies the problem significantly,
but at the price of rendering the total energy a non-conserved
quantity
on the MC-TDDFT level. Furthermore, 
the density dependence
of existing effective interactions
requires employing a density prescription,
which can be seen as an additional parameter of the model.

A technical discussion was supplemented with
numerical examples, focusing on the issues of convergence,
treatment of linearly dependent bases, energy conservation,
and prescriptions for the density-dependent part of an
effective interaction. To start with, we demonstrated
the convergence
of static and dynamic aspects of the new TDDFT solver,
based on the finite element method. The time evolution
of the MC-TDDFT
state was shown to be unitary and well converged for a wide range
of time steps and spatial meshes. Generally,
finer spatial meshes require smaller time steps.
Similarly, MC-TDDFT calculations require finer meshes 
than those used in TDDFT to 
achieve a comparable level of convergence. 
Linear dependencies in the
basis need to be carefully treated, since including
too small norm eigenvalues causes numerical instabilities,
while excluding too large eigenvalues removes a part
of physical information. The non-conservation of the
MC-TDDFT energy is a combined effect
of the kinetic, nuclear, and Coulomb components,
and it is at the order of $0.5\%$ in the 
considered example.
Finally, the two versions of the average density
prescription discussed in this work
yield only minor differences
in the collective dynamics.

The very recent
implementations of the MC-TDDFT framework
in real nuclei \cite{marevic2023,li2023,li2023b},
based on the pioneering
work by Reinhard and collaborators 
\cite{reinhard1983,reinhard1987} and
following the toy-model
study of \cite{regnier2019}, 
have
demonstrated the predictive
power
of the model. Further developments
of the theoretical framework and computational
methods are expected to render the model
applicable to a wider range of
nuclear phenomena in the near future. 
Future studies should also aim at elucidating the relation and difference in performance between the MC-TDDFT and the conventional TDGCM framework.
 
\appendix

\section{Local transition densities}
\label{Sec:TransitionDensities}

The spin expansion of the
non-local transition density
[Eq.~\eqref{eq:NonLocalDensity}]
for isospin $\tau$ reads
\begin{equation}
\begin{split}
\rho_{\bm{q}\bm{q'}}^{(\tau)}(\bm{r}\sigma,\bm{r'}\sigma';t)&= \frac{1}{2}\rho_{\bm{q}\bm{q'}}^{(\tau)}(\bm{r},\bm{r'};t) \delta_{\sigma \sigma'} \\
&+ \frac{1}{2} \sum_\mu \braket{\sigma|\hat{\sigma}_\mu 
|\sigma'}s_{\bm{q}\bm{q'}, \mu}^{(\tau)} (\bm{r},\bm{r'};t),
\end{split}
\end{equation}
where $\rho_{\bm{q}\bm{q'}}^{(\tau)}(\bm{r},\bm{r'};t)$
is the non-local one-body transition particle
density,
\begin{equation}
\rho_{\bm{q}\bm{q'}}^{(\tau)}(\bm{r},\bm{r'};t) = \sum_\sigma
\rho_{\bm{q}\bm{q'}}^{(\tau)}(\bm{r}\sigma,\bm{r'}\sigma;t),
\end{equation}
$s_{\bm{q}\bm{q'}, \mu}^{(\tau)} (\bm{r},\bm{r'};t)$ is the
$\mu$-th component of the non-local
one-body transition spin density,
\begin{equation}
s_{\bm{q}\bm{q'}, \mu}^{(\tau)} (\bm{r},\bm{r'};t) = 
\sum_{\sigma \sigma'} \rho_{\bm{q}\bm{q'}}^{(\tau)}(\bm{r}\sigma,\bm{r'}\sigma';t)
\braket{\sigma'|\hat{\sigma}_\mu | \sigma},
\end{equation}
and $\hat{\sigma}_\mu$
are the Pauli operators.
The local variants of
the particle density $\rho_{\bm{q} \bm{q'}}^{(\tau)}(\bm{r};t)$,
spin density $\bm{s}_{\bm{q} \bm{q'}}^{(\tau)}(\bm{r};t)$,
kinetic density
$\tau_{\bm{q} \bm{q'}}^{(\tau)}(\bm{r};t)$,
current density
$\bm{j}_{\bm{q} \bm{q'}}^{(\tau)}(\bm{r};t)$,
spin-current pseudotensor density
$J_{\bm{q} \bm{q'}, \mu \nu}^{(\tau)}(\bm{r};t)$,
and spin-orbit current vector density
${\rm{\bm{J}}}_{\bm{q} \bm{q'}}^{(\tau)}(\bm{r};t)$
read
\begin{subequations}
\begin{align}
\rho_{\bm{q} \bm{q'}}^{(\tau)}(\bm{r};t) &= 
\rho_{\bm{q} \bm{q'}}^{(\tau)}(\bm{r}, \bm{r};t),\\
s_{\bm{q} \bm{q'},\mu}^{(\tau)}(\bm{r};t) &= 
s_{\bm{q} \bm{q'},\mu}^{(\tau)}(\bm{r}, \bm{r};t),\\
\tau_{\bm{q} \bm{q'}}^{(\tau)}(\bm{r};t) &=
\nabla \cdot \nabla' \rho_{\bm{q} \bm{q'}}^{(\tau)}
(\bm{r},\bm{r'};t)\vert_{\bm{r'}=\bm{r}}, \\
j_{\bm{q} \bm{q'}, \mu}^{(\tau)}(\bm{r};t) &=
\frac{1}{2\di}(\nabla_\mu - \nabla'_\mu)\rho_{\bm{q} \bm{q'}}^{(\tau)}
(\bm{r},\bm{r'};t)\vert_{\bm{r'}=\bm{r}}, \\
J_{\bm{q} \bm{q'},\mu \nu}^{(\tau)}(\bm{r};t) &=
\frac{1}{2\di}(\nabla_\mu - \nabla'_\mu)
s_{\bm{q} \bm{q'},\nu}^{(\tau)}(\bm{r},\bm{r'};t)\vert_{\bm{r'}=\bm{r}}, \\
{\rm{J}}_{\bm{q} \bm{q'}, \lambda}^{(\tau)}(\bm{r};t) &= \sum_{\mu \nu} \epsilon_{\lambda \mu \nu } J_{\bm{q} \bm{q'},\mu \nu}^{(\tau)}(\bm{r};t).
\end{align}
\end{subequations}
In the following paragraph, the explicit dependence on time and isospin 
is omitted for compactness.

For Slater generating states, the coordinate
space representation
of the non-local
transition density (for either neutrons
or protons)
can be written as
\begin{equation}
\rho_{\bm{q} \bm{q'}}(\bm{r}\sigma,\bm{r'}\sigma')
= \sum_{kl}\varphi_k^{\bm{q'}}(\bm{r}\sigma)
\Big[M_{\bm{q} \bm{q'}}^{-1}\Big]_{kl}\varphi_l^{\bm{q}*}
(\bm{r'}\sigma'),
\end{equation}
where $\Big[M_{\bm{q} \bm{q'}}^{-1}(t)\Big]_{kl}$ are (generally complex)
elements of the inverted matrix of single-particle
overlaps [Eq.~\eqref{eq:OverlapMatrix}].
Given the decomposition \eqref{eq:SpDecomposition}, 
the local transition particle density
reads
\begin{equation}
\rho_{\bm{q}\bm{q'}}(\bm{r}) = \sum_{kl} \Big[M_{\bm{q}\bm{q'}}^{-1} \Big]_{kl} \Big[
\rho_{\bm{q}\bm{q'}}^{R}(\bm{r})+\di\rho_{\bm{q}\bm{q'}}^{I}(\bm{r}) \Big]_{kl}
\label{eq:ParticleDensity}
\end{equation}
with
\begin{subequations}
\begin{align}
\Big[\rho_{\bm{q}\bm{q'}}^{R}(\bm{r})\Big]_{kl} & =
\sum_{\alpha} \varphi_{k,\alpha}^{\bm{q'}}
(\bm{r})
\varphi_{l,\alpha}^{\bm{q}}(\bm{r}),  \\
\Big[\rho_{\bm{q}\bm{q'}}^{I}(\bm{r})\Big]_{kl} & =
\varphi_{k,1}^{\bm{q'}}(\bm{r})
\varphi_{l,0}^{\bm{q}}(\bm{r})
-\varphi_{k,0}^{\bm{q'}}(\bm{r})
\varphi_{l,1}^{\bm{q}}(\bm{r})
\nonumber
\\ & +\varphi_{k,3}^{\bm{q'}}(\bm{r})
\varphi_{l,2}^{\bm{q}}(\bm{r})
-\varphi_{k,2}^{\bm{q'}} (\bm{r})
\varphi_{l,3}^{\bm{q}}(\bm{r}),
\end{align}
\end{subequations}
for $\alpha = 0, 1, 2, 3$.
Similarly, the local transition kinetic density
reads
\begin{equation}
\tau_{\bm{q}\bm{q'}}(\bm{r}) = \sum_{kl}
\Big[M_{\bm{q}\bm{q'}}^{-1}\Big]_{kl}
\Big[\tau_{\bm{q}\bm{q'}}^{R}(\bm{r})+\di\tau_{\bm{q}\bm{q'}}^{I}(\bm{r})\Big]_{kl}, 
\label{eq:KineticDensity}
\end{equation}
with
\begin{subequations}
\begin{align}
\Big[\tau_{\bm{q}\bm{q'}}^{R}(\bm{r})\Big]_{kl} & = \sum_{\alpha}
\big(\nabla \varphi_{k,\alpha}^{\bm{q'}}(\bm{r})\big)\big(\nabla\varphi_{l,\alpha}^{\bm{q}}(\bm{r})\big),
\\ 
\Big[\tau_{\bm{q}\bm{q'}}^{I}(\bm{r})\Big]_{kl} & =
\big(\nabla \varphi_{k,1}^{\bm{q'}}(\bm{r})\big) \big(\nabla \varphi_{l,0}^{\bm{q}}(\bm{r})\big)
\nonumber \\ &
- \big(\nabla \varphi_{k,0}^{\bm{q'}}(\bm{r})\big)\big(\nabla \varphi_{l,1}^{\bm{q}}(\bm{r})\big)
\nonumber \\ & + \big(\nabla \varphi_{k,3}^{\bm{q'}}(\bm{r})\big) \big(\nabla \varphi_{l,2}^{\bm{q}}(\bm{r})\big) \nonumber  \\
& - \big(\nabla \varphi_{k,2}^{\bm{q'}}(\bm{r})\big)\big(\nabla \varphi_{l,3}^{\bm{q}}(\bm{r})\big),
\end{align}
\end{subequations}
with
\begin{equation}
\big(\nabla \varphi_{k,\alpha}^{\bm{q'}}(\bm{r})\big)\big(\nabla\varphi_{l,\beta}^{\bm{q}}(\bm{r})\big) =
\sum_{\mu}\big(\partial_\mu \varphi_{k,\alpha}^{\bm{q'}}(\bm{r})\big) \big(\partial_\mu \varphi_{l,\beta}^{\bm{q}}(\bm{r})\big)
\end{equation}
and $\mu = x, y, z$.
Furthermore, the $\mu$-th component of the
local transition current density reads
\begin{equation}
j_{\bm{q}\bm{q'}}^{\mu}(\bm{r}) =
\frac{1}{2}\sum_{kl}\Big[M_{\bm{q}\bm{q'}}^{-1}\Big]_{kl} 
\Big[j_{\bm{q}\bm{q'}}^{\mu,R}(\bm{r})+\di j_{\bm{q} \bm{q'}}^{\mu,I}(\bm{r})\Big]_{kl},
\end{equation}
with
\begin{subequations}
\begin{align}
\Big[j_{\bm{q}\bm{q'}}^{\mu,R}(\bm{r})\Big]_{kl} & = 
\big(\partial_\mu \varphi_{k,1}^{\bm{q'}}(\bm{r})\big)
\varphi_{l,0}^{\bm{q}}(\bm{r})
-\big(\partial_\mu \varphi_{k,0}^{\bm{q'}}(\bm{r})\big)
\varphi_{l,1}^{\bm{q}}(\bm{r}) \nonumber \\
& + \big(\partial_\mu \varphi_{k,3}^{\bm{q'}}(\bm{r})\big)
\varphi_{l,2}^{\bm{q}}(\bm{r}) -
\big(\partial_\mu \varphi_{k,2}^{\bm{q'}}(\bm{r})\big)
\varphi_{l,3}^{\bm{q}}(\bm{r}) \nonumber \\
& - \varphi_{k,1}^{\bm{q'}}(\bm{r})\big(\partial_\mu \varphi_{l,0}^{\bm{q}}(\bm{r})\big)
+ \varphi_{k,0}^{\bm{q'}}(\bm{r})\big(\partial_\mu \varphi_{l,1}^{\bm{q}}(\bm{r})\big) \nonumber \\
& - \varphi_{k,3}^{\bm{q'}}(\bm{r})\big(\partial_\mu \varphi_{l,2}^{\bm{q}}(\bm{r})\big)
+ \varphi_{k,2}^{\bm{q'}}(\bm{r})\big(\partial_\mu \varphi_{l,3}^{\bm{q}}(\bm{r})\big), \\
\Big[j_{\bm{q}\bm{q'}}^{\mu,I}(\bm{r})\Big]_{kl} & = 
\sum_{\alpha}
\varphi_{k,\alpha}^{\bm{q'}}(\bm{r})
\big(\partial_\mu \varphi_{l,\alpha}^{\bm{q}}
(\bm{r})\big) \nonumber \\
& -  \sum_{\alpha}\big(\partial_\mu \varphi_{k,\alpha}^{\bm{q'}}(\bm{r})\big)
\varphi_{l,\alpha}^{\bm{q}}(\bm{r}),
\end{align}
\end{subequations}
The components of the local transition
spin density then read
\begin{equation}
s_{\bm{q}\bm{q'}}^{\mu}(\bm{r}) = 
\sum_{kl} \Big[M_{\bm{q}\bm{q'}}^{-1}\Big]_{kl}
\Big[s_{\bm{q}\bm{q'}}^{\mu,R}(\bm{r})
+\di s_{\bm{q}\bm{q'}}^{\mu,I}(\bm{r})\Big]_{kl}, 
\end{equation}
with
\begin{subequations}
\begin{align}
\Big[s_{\bm{q}\bm{q'}}^{x,R}(\bm{r})\Big]_{kl} &= 
\varphi_{k,0}^{\bm{q'}}(\bm{r})\varphi_{l,2}^{\bm{q}}(\bm{r}) +
\varphi_{k,1}^{\bm{q'}}(\bm{r})
\varphi_{l,3}^{\bm{q}}(\bm{r}) \nonumber \\
& + \varphi_{k,2}^{\bm{q'}}(\bm{r})\varphi_{l,0}^{\bm{q}}(\bm{r}) +
\varphi_{k,3}^{\bm{q'}}(\bm{r})
\varphi_{l,1}^{\bm{q}}(\bm{r}), \\
\Big[s_{\bm{q}\bm{q'}}^{x,I}(\bm{r})\Big]_{kl} &= 
\varphi_{k,1}^{\bm{q'}}(\bm{r})
\varphi_{l,2}^{\bm{q}}(\bm{r}) -
\varphi_{k,2}^{\bm{q'}}(\bm{r})
\varphi_{l,1}^{\bm{q}}(\bm{r}) \nonumber \\
& + \varphi_{k,3}^{\bm{q'}}(\bm{r})\varphi_{l,0}^{\bm{q}}(\bm{r}) -
\varphi_{k,0}^{\bm{q'}}(\bm{r})
\varphi_{l,3}^{\bm{q}}(\bm{r}), \\
\Big[s_{\bm{q}\bm{q'}}^{y,R}(\bm{r})\Big]_{kl} &= 
\varphi_{k,0}^{\bm{q'}}(\bm{r})
\varphi_{l,3}^{\bm{q}}(\bm{r}) +
\varphi_{k,3}^{\bm{q'}}(\bm{r})
\varphi_{l,0}^{\bm{q}}(\bm{r}) \nonumber \\
& - \varphi_{k,1}^{\bm{q'}}(\bm{r})
\varphi_{l,2}^{\bm{q}}(\bm{r}) -
\varphi_{k,2}^{\bm{q'}}(\bm{r})
\varphi_{l,1}^{\bm{q}}(\bm{r}), \\
\Big[s_{\bm{q}\bm{q'}}^{y,I}(\bm{r})\Big]_{kl} &= 
\varphi_{k,0}^{\bm{q'}}(\bm{r})
\varphi_{l,2}^{\bm{q}}(\bm{r}) -
\varphi_{k,2}^{\bm{q'}}(\bm{r})
\varphi_{l,0}^{\bm{q}}(\bm{r}) \nonumber \\
& + \varphi_{k,1}^{\bm{q'}}(\bm{r})
\varphi_{l,3}^{\bm{q}}(\bm{r}) -
\varphi_{k,3}^{\bm{q'}}(\bm{r})
\varphi_{l,1}^{\bm{q}}(\bm{r}),
\\
\Big[s_{\bm{q}\bm{q'}}^{z,R}(\bm{r})\Big]_{kl} &= 
\varphi_{k,0}^{\bm{q'}}(\bm{r})
\varphi_{l,0}^{\bm{q}}(\bm{r}) +
\varphi_{k,1}^{\bm{q'}}(\bm{r})
\varphi_{l,1}^{\bm{q}}(\bm{r}) \nonumber \\
& - \varphi_{k,2}^{\bm{q'}}(\bm{r})\varphi_{l,2}^{\bm{q}}(\bm{r}) -
\varphi_{k,3}^{\bm{q'}}(\bm{r})
\varphi_{l,3}^{\bm{q}}(\bm{r}), \\
\Big[s_{\bm{q}\bm{q'}}^{z,I}(\bm{r})\Big]_{kl} &= 
\varphi_{k,1}^{\bm{q'}}(\bm{r})
\varphi_{l,0}^{\bm{q}}(\bm{r}) -
\varphi_{k,0}^{\bm{q'}}(\bm{r})
\varphi_{l,1}^{\bm{q}}(\bm{r}) \nonumber \\
& + \varphi_{k,2}^{\bm{q'}}(\bm{r})\varphi_{l,3}^{\bm{q}}(\bm{r}) -
\varphi_{k,3}^{\bm{q'}}(\bm{r})
\varphi_{l,2}^{\bm{q}}(\bm{r}).
\end{align}
\end{subequations}
Finally, the
components of the spin-current pseudotensor density read
\begin{equation}
J_{\bm{q}\bm{q'}}^{\mu \nu}(\bm{r}) = 
\frac{1}{2}
\sum_{kl}\Big[M_{\bm{q}\bm{q'}}^{-1}\Big]_{kl}
\Big[J_{\bm{q}\bm{q'}}^{\mu \nu, R}(\bm{r})
+\di J_{\bm{q}\bm{q'}}^{\mu \nu, I}(\bm{r})\Big]_{kl}
\end{equation}
with
\begin{subequations}
\begin{align}
\Big[J_{\bm{q}\bm{q'}}^{\mu x, R}(\bm{r})\Big]_{kl} &=
\big(\partial_\mu \varphi_{k,1}^{\bm{q'}}(\bm{r})\big)
\varphi_{l,2}^{\bm{q}}(\bm{r}) - 
\varphi_{k,1}^{\bm{q'}}(\bm{r})\big(\partial_\mu 
\varphi_{l,2}^{\bm{q}}(\bm{r})\big) \nonumber 
\\
& - \big(\partial_\mu \varphi_{k,0}^{\bm{q'}}(\bm{r})\big)
\varphi_{l,3}^{\bm{q}}(\bm{r})
+ \varphi_{k,0}^{\bm{q'}}(\bm{r})
\big(\partial_\mu \varphi_{l,3}^{\bm{q}}(\bm{r})\big) \nonumber \\
& + \big(\partial_\mu \varphi_{k,3}^{\bm{q'}}(\bm{r})\big)
\varphi_{l,0}^{\bm{q}}(\bm{r})
- \varphi_{k,3}^{\bm{q'}}(\bm{r})
\big(\partial_\mu \varphi_{l,0}^{\bm{q}}(\bm{r})\big) \nonumber \\
& - \big(\partial_\mu \varphi_{k,2}^{\bm{q'}}(\bm{r})\big)
\varphi_{l,1}^{\bm{q}}(\bm{r})
+ \varphi_{k,2}^{\bm{q'}}(\bm{r})
\big(\partial_\mu \varphi_{l,1}^{\bm{q}}(\bm{r})\big), \\
\Big[J_{\bm{q}\bm{q'}}^{\mu x, I}(\bm{r})\Big]_{kl} &= 
\varphi_{k,0}^{\bm{q'}}(\bm{r})
\big(\partial_\mu 
\varphi_{l,2}^{\bm{q}}(\bm{r})\big)
-
\big(\partial_\mu \varphi_{k,0}^{\bm{q'}}(\bm{r})\big)
\varphi_{l,2}^{\bm{q}}(\bm{r})  \nonumber \\
& + \varphi_{k,1}^{\bm{q'}}(\bm{r})
\big(\partial_\mu \varphi_{l,3}^{\bm{q}}(\bm{r})\big)
- \big(\partial_\mu \varphi_{k,1}^{\bm{q'}}(\bm{r})\big)
\varphi_{l,3}^{\bm{q}}(\bm{r})
\nonumber \\
& + \varphi_{k,2}^{\bm{q'}}(\bm{r})
\big(\partial_\mu \varphi_{l,0}^{\bm{q}}(\bm{r})\big)
- \big(\partial_\mu \varphi_{k,2}^{\bm{q'}}(\bm{r})\big)
\varphi_{l,0}^{\bm{q}}(\bm{r})
\nonumber \\
& + \varphi_{k,3}^{\bm{q'}}(\bm{r})
\big(\partial_\mu \varphi_{l,1}^{\bm{q}}(\bm{r})\big)
- \big(\partial_\mu \varphi_{k,3}^{\bm{q'}}(\bm{r})\big)
\varphi_{l,1}^{\bm{q}}(\bm{r}), \\
\Big[J_{\bm{q}\bm{q'}}^{\mu y,R}(\bm{r})\Big]_{kl} &=
\big(\partial_\mu \varphi_{k,0}^{\bm{q'}}(\bm{r})\big)
\varphi_{l,2}^{\bm{q}}(\bm{r}) - 
\varphi_{k,0}^{\bm{q'}}(\bm{r})
\big(\partial_\mu 
\varphi_{l,2}^{\bm{q}}(\bm{r})\big) \nonumber \\
& + \big(\partial_\mu \varphi_{k,1}^{\bm{q'}}(\bm{r})\big)
\varphi_{l,3}^{\bm{q}}(\bm{r})
- \varphi_{k,1}^{\bm{q'}}(\bm{r})
\big(\partial_\mu \varphi_{l,3}^{\bm{q}}(\bm{r})\big) \nonumber \\
& - \big(\partial_\mu \varphi_{k,2}^{\bm{q'}}(\bm{r})
\big)
\varphi_{l,0}^{\bm{q}}(\bm{r})
+ \varphi_{k,2}^{\bm{q'}}(\bm{r})
\big(\partial_\mu \varphi_{l,0}^{\bm{q}}(\bm{r})
\big) \nonumber \\
& - \big(\partial_\mu \varphi_{k,3}^{\bm{q'}}
(\bm{r})\big)
\varphi_{l,1}^{\bm{q}}(\bm{r})
+ \varphi_{k,3}^{\bm{q'}}(\bm{r})
\big(\partial_\mu \varphi_{l,1}^{\bm{q}}
(\bm{r})\big), \\
\Big[J_{\bm{q}\bm{q'}}^{\mu y,I}(\bm{r})\Big]_{kl} &= 
\big(\partial_\mu \varphi_{k,1}^{\bm{q'}}(\bm{r})\big)
\varphi_{l,2}^{\bm{q}}(\bm{r}) - 
\varphi_{k,1}^{\bm{q'}}(\bm{r})
\big(\partial_\mu 
\varphi_{l,2}^{\bm{q}}(\bm{r})\big) \nonumber \\
& - \big(\partial_\mu \varphi_{k,0}^{\bm{q'}}
(\bm{r})\big)
\varphi_{l,3}^{\bm{q}}(\bm{r})
+ \varphi_{k,0}^{\bm{q'}}(\bm{r})
\big(\partial_\mu \varphi_{l,3}^{\bm{q}}(\bm{r})\big) \nonumber \\
& + \big(\partial_\mu \varphi_{k,2}^{\bm{q'}}
(\bm{r})\big)
\varphi_{l,1}^{\bm{q}}(\bm{r})
- \varphi_{k,2}^{\bm{q'}}(\bm{r})
\big(\partial_\mu \varphi_{l,1}^{\bm{q}}(\bm{r})\big) \nonumber \\
& - \big(\partial_\mu \varphi_{k,3}^{\bm{q'}}(\bm{r})\big)
\varphi_{l,0}^{\bm{q}}(\bm{r})
+ \varphi_{k,3}^{\bm{q'}}(\bm{r})
\big(\partial_\mu \varphi_{l,0}^{\bm{q}}(\bm{r})\big), \\
\Big[J_{\bm{q}\bm{q'}}^{\mu z, R}(\bm{r})\Big]_{kl} &=
\big(\partial_\mu \varphi_{k,1}^{\bm{q'}}(\bm{r})\big)
\varphi_{l,0}^{\bm{q}}(\bm{r}) - 
\varphi_{k,1}^{\bm{q'}}(\bm{r})
\big(\partial_\mu 
\varphi_{l,0}^{\bm{q}}(\bm{r})\big) \nonumber \\
& - \big(\partial_\mu \varphi_{k,0}^{\bm{q'}}
(\bm{r})\big)
\varphi_{l,1}^{\bm{q}}(\bm{r})
+ \varphi_{k,0}^{\bm{q'}}(\bm{r})
\big(\partial_\mu \varphi_{l,1}^{\bm{q}}(\bm{r})\big) \nonumber \\
& + \big(\partial_\mu \varphi_{k,2}^{\bm{q'}}(\bm{r})\big)
\varphi_{l,3}^{\bm{q}}(\bm{r})
- \varphi_{k,2}^{\bm{q'}}(\bm{r})
\big(\partial_\mu \varphi_{l,3}^{\bm{q}}(\bm{r})
\big) \nonumber \\
& - \big(\partial_\mu \varphi_{k,3}^{\bm{q'}}(\bm{r})\big)
\varphi_{l,2}^{\bm{q}}(\bm{r})
+ \varphi_{k,3}^{\bm{q'}}(\bm{r})
\big(\partial_\mu \varphi_{l,2}^{\bm{q}}(\bm{r})\big), \\
\Big[J_{\bm{q}\bm{q'}}^{\mu z, I}(\bm{r})\Big]_{kl} &=  
\varphi_{k,0}^{\bm{q'}}(\bm{r})
\big(\partial_\mu 
\varphi_{l,0}^{\bm{q}}(\bm{r})\big) 
-  
\big(\partial_\mu \varphi_{k,0}^{\bm{q'}}(\bm{r})
\big)
\varphi_{l,0}^{\bm{q}}(\bm{r})
\nonumber \\
& + \varphi_{k,1}^{\bm{q'}}(\bm{r})
\big(\partial_\mu \varphi_{l,1}^{\bm{q}}(\bm{r})
\big) 
- \big(\partial_\mu \varphi_{k,1}^{\bm{q'}}(\bm{r})
\big)
\varphi_{l,1}^{\bm{q}}(\bm{r})
\nonumber \\
&
- \varphi_{k,2}^{\bm{q'}}(\bm{r})
\big(\partial_\mu \varphi_{l,2}^{\bm{q}}(\bm{r})\big) 
+ \big(\partial_\mu \varphi_{k,2}^{\bm{q'}}(\bm{r})
\big)
\varphi_{l,2}^{\bm{q}}(\bm{r})
\nonumber \\
& - \varphi_{k,3}^{\bm{q'}}(\bm{r})
\big(\partial_\mu \varphi_{l,3}^{\bm{q}}(\bm{r})\big)
+ \big(\partial_\mu \varphi_{k,3}^{\bm{q'}}(\bm{r})\big)
\varphi_{l,3}^{\bm{q}}(\bm{r}).
\end{align}
\end{subequations}

\section{Coupling constants}
\label{sec:CouplingConstants}

The coupling constants appearing in the
Skyrme energy density \eqref{eq:skyrmekernel}
read

\begin{subequations}
\begin{align}
B_1 & = \frac{1}{2}t_0(1+\frac{1}{2}x_0) ,\\
B_2 & = - \frac{1}{2}t_0(\frac{1}{2} + x_0), \\
B_3 & =
\frac{1}{4}
\big(t_1 (1 + \frac{1}{2}x_1)
+ t_2 (1 + \frac{1}{2}x_2)\big), \\
B_4 & = -\frac{1}{4}
\big(t_1(\frac{1}{2} + x_1) - t_2(\frac{1}{2} + x_2)\big),
\\
B_5 & = -\frac{1}{16}
\big(3t_1(1+\frac{1}{2}x_1)
- t_2(1+\frac{1}{2}x_2)\big),
\\
B_6 & = 
\frac{1}{16}
\big(3t_1 (\frac{1}{2}+ x_1)
+ t_2(\frac{1}{2} + x_2)\big),
\\
B_7 & = 
\frac{1}{12}t_3(1+\frac{1}{2}x_3),
\\
B_8 & = 
- \frac{1}{12}t_3(\frac{1}{2} + x_3) ,\\
B_9 & = -\frac{1}{2}W,
\\
B_{10} & = \frac{1}{4}t_0x_0,
\\
B_{11} & = -\frac{1}{4}t_0 ,\\
B_{12} & = \frac{1}{24}t_3x_3, \\
B_{13} & = -\frac{1}{24}t_3. 
\end{align}
\end{subequations}
Parameters $t_i$, $x_i$ ($i = 0, 1, 2, 3$), $W$, and
$\alpha$
are the standard parameters
of the Skyrme pseudopotential \cite{schunck2019,bonche1987}.

\section{Variance of a one-body
operator in a 
normalized MC-TDDFT
state}
\label{sec:variance}

To evaluate the variance of
Eq.~\eqref{eq:variance},
we need an expectation
value of the $\hat{O}^2$
operator in a normalized
MC-TDDFT state.
We start from
\begin{equation}
\braket{\Psi(t)|\hat{O}^2|\Psi(t)}
= \int_{\bm{q} \bm{q'}}  
\! \! \,d\bm{q} \,d\bm{q'}
g^*_{\bm{q}}(t) {\mathcal{O}^2}^c_{\bm{q}\bm{q'}}(t)
g_{\bm{q'}}(t).
\end{equation}
Again, the collective kernel
of the $\hat{O}^2$ operator
is calculated from \eqref{eq:collective_operators}
and the corresponding usual kernel
follows
from the generalized Wick
theorem,
\begin{equation}
\begin{split}
\mathcal{O}^2_{\bm{q} \bm{q'}}(t)
= \mathcal{N}_{\bm{q} \bm{q'}}(t)
\Big[&{\rm{Tr}}^2\big(O\rho^{\bm{q}\bm{q'}}(t)\big) \\
+ & {\rm{Tr}}\big(O\rho^{\bm{q}\bm{q'}}(t)O(1-\rho^{\bm{q}\bm{q'}}(t)\big)\Big].
\end{split}
\end{equation}
Here, 
\begin{equation}
{\rm{Tr}}\big(O\rho^{\bm{q}\bm{q'}}(t)\big)
= \int \,d^3 \bm{r}
O(\bm{r}) \rho_{\bm{q} \bm{q'}}(\bm{r}; t).
\end{equation}
Furthermore, the second trace
corresponds to the sum of two
terms, 
\begin{equation}
{\rm{Tr}}\big(O\rho^{\bm{q}\bm{q'}}(t)O(1-\rho^{\bm{q}\bm{q'}}(t)\big) = 
C_1^{\bm{q} \bm{q'}}(t)
+ C_2^{\bm{q} \bm{q'}}(t).
\end{equation}
The first term reads
\begin{equation}
\begin{split}
C_1^{\bm{q} \bm{q'}}(t) & = 
\int \,d^3 \bm{r}
O^2(\bm{r}) \sum_{kl}
\Big[M_{\bm{q} \bm{q'}}^{-1}(t)\Big]_{kl}
\\ & \times \biggl\{ A_{kl}^{\bm{q}\bm{q'}}(\bm{r};t)
+ \di B_{kl}^{\bm{q}\bm{q'}}(\bm{r};t) \biggl\},
\end{split}
\end{equation}
with
\begin{equation}
A_{kl}^{\bm{q}\bm{q'}}(\bm{r};t) = \sum_{\alpha}
\varphi_{k,\alpha}^{\bm{q'}}(\bm{r};t) \varphi_{l,\alpha}^{\bm{q}}(\bm{r};t)
\end{equation}
and
\begin{equation}
\begin{split}
B_{kl}^{\bm{q}\bm{q'}}(\bm{r};t) &= 
\varphi_{k,1}^{\bm{q'}}(\bm{r};t) \varphi_{l,0}^{\bm{q}}(\bm{r};t)
\\ & - \varphi_{k,0}^{\bm{q'}}(\bm{r};t) \varphi_{l,1}^{\bm{q}}(\bm{r};t) \\ & +
\varphi_{k,3}^{\bm{q'}}(\bm{r};t) \varphi_{l,2}^{\bm{q}}(\bm{r};t)
\\ & - \varphi_{k,2}^{\bm{q'}}(\bm{r};t) \varphi_{l,3}^{\bm{q}}(\bm{r};t).
\end{split}
\end{equation}
The second term reads
\begin{equation}
\begin{split}
C_2^{\bm{q} \bm{q'}}(t)
& = \int \,d^3 \bm{r}
\int \,d^3 \bm{r'}
O(\bm{r}) 
O(\bm{r'})
\\ & \times \sum_{klmn}
\Big[M_{\bm{q} \bm{q'}}^{-1}(t)\Big]_{kl}
\Big[M_{\bm{q} \bm{q'}}^{-1}(t)\Big]_{mn} \\
& \times 
\biggl\{ A_{kn}^{\bm{q}\bm{q'}}(\bm{r};t)
+ \di B_{kn}^{\bm{q}\bm{q'}}(\bm{r};t) \biggl\}
\\ & \times
\biggl\{ A_{ml}^{\bm{q}\bm{q'}}(\bm{r'};t)
+ \di B_{ml}^{\bm{q}\bm{q'}}(\bm{r'};t) \biggl\}.
\end{split}
\end{equation}

\begin{acknowledgements}
This work was supported in part by CNRS through 
the AIQI-IN2P3 funding.
P. M. would like to express his gratitude to
CEA and IJCLab for their warm hospitality 
during work on this project.
\end{acknowledgements}

\bibliographystyle{elsarticle-num}
\bibliography{bibliography}  

\end{document}